\documentclass[prb,superscriptaddress,amsmath,amssymb,floatfix,showpacs,twocolumn]{revtex4-1}
\usepackage{amsfonts,amsmath,amssymb}
\usepackage{epsfig}
\usepackage{times}
\usepackage{graphicx}
\usepackage{bm}
\usepackage{color}
\usepackage{dcolumn}
\usepackage{array}
\usepackage{float}
\usepackage{soul}

\usepackage{feynmp-auto}
\newcommand\xin{x_{\text{in}}}
\newcommand\xout{x_{\text{out}}}
\newcommand\wc{\text{\emph{w}}_\mathfrak{C}}
\newcommand\wcp{\text{\emph{w}}_\mathfrak{C'}}
\newcommand\cc{\mathfrak{C}}

\begin{document}
\unitlength = 1mm

\title{Determinant Monte Carlo algorithms for dynamical quantities
       in fermionic systems}

\author{Alice Moutenet}
\affiliation{Centre de Physique Th\'eorique, Ecole Polytechnique, CNRS, Universit\'e Paris-Saclay, 91128 Palaiseau, France}
\affiliation{Coll\`ege de France, 11 place Marcelin Berthelot, 75005 Paris, France}
\author{Wei Wu}
\affiliation{Centre de Physique Th\'eorique, Ecole Polytechnique, CNRS, Universit\'e Paris-Saclay, 91128 Palaiseau, France}
\affiliation{Coll\`ege de France, 11 place Marcelin Berthelot, 75005 Paris, France}
\author{Michel Ferrero}
\affiliation{Centre de Physique Th\'eorique, Ecole Polytechnique, CNRS, Universit\'e Paris-Saclay, 91128 Palaiseau, France}
\affiliation{Coll\`ege de France, 11 place Marcelin Berthelot, 75005 Paris, France}

\date{\today}

\begin{abstract}

We introduce and compare three different Monte Carlo determinantal algorithms
that allow one to compute dynamical quantities, such as the self-energy, of
fermionic systems in their thermodynamic limit.
We show that the most efficient approach expresses the sum of a factorial
number of one-particle-irreducible diagrams as a recursive sum of determinants
with exponential complexity. 
By comparing results for the two-dimensional Hubbard model with those obtained
from state-of-the-art diagrammatic Monte Carlo, we show that we can reach
higher perturbation orders and greater accuracy for the same computational
effort.

\end{abstract}

\maketitle


\section{Introduction}

Perturbation expansions are at the heart of many important developments in
many-body physics.
They appear both in the construction of new theoretical frameworks and in the
design of numerical algorithms that have greatly contributed to push further
our understanding of interacting quantum systems.

Continuous-time quantum Monte Carlo algorithms~\cite{gull_rmp_2011} such as
CT-INT,~\cite{rubtsov_2004, rubtsov_prb_2005} CT-AUX~\cite{gull_epl_2008} or
CT-HYB~\cite{werner_prl_2006, werner_prb_2006} are examples of such algorithms.
They have been a breakthrough in finding solutions of quantum impurity problems
and have opened a new realm for the development of extensions of dynamical
mean-field theory.~\cite{antoine_rmp_1996, kotliar_rmp_2006, hettler_prb_1998,
lichtenstein_prb_2000, kotliar_prl_2001, maier_rmp_2005}

One of the reasons for the success of these algorithms is that they are based
on a perturbation expansion of the partition function $Z$. The contributions to
$Z$ can be reorganized  into determinants that effectively sum a factorial
number of perturbation diagrams. As a result, large perturbation orders can be
computed and, for smaller clusters, the strong-coupling, low-temperature regime
can be addressed. These methods are however limited by the number of sites that
can be treated in the auxiliary quantum impurity cluster. For large clusters
the fermionic sign problem~\cite{troyer_prl_2005} becomes very severe as
temperature is decreased or interaction increased~\cite{leblanc_prx_2015} and
it is very difficult to extrapolate the solution of the infinite size system
from a limited number of small clusters.

An alternative and complementary approach is to investigate quantum systems
directly in their thermodynamic limit, as in the
DiagMC~\cite{prokofev_prl_1998, prokofev_prl_2007, kozik_epl_2010,
vanhoucke_2010, burovski_prb_2004} algorithm that has also benefited from great
advances. With this approach, controlled results have been obtained, e.g. for the
normal phase of the unitary Fermi gas,~\cite{vanhoucke_nature_2012,
burovski_prl_2006} the ground-state phase diagram of the Hubbard model in the
weak-coupling regime away from half-filling,~\cite{gukelberger_prb_2015,
deng_epl_2015, fedor_prb_2016, fedor_prb_2017} and even in parts
of its phase diagram where a pseudogap has already formed.~\cite{wei_prb_2017}

In this method, the perturbation series is written directly for the physical
quantity of interest, for example the self-energy. Contributions to the series
are given by individual perturbation Feynman diagrams (one-particle irreducible
ones for the self-energy) that are sampled stochastically.
While the sign alternation between individual diagrams is a necessary condition
for the convergence of the series, it introduces a fermionic sign problem that
makes it difficult to precisely compute high-order coefficients of the series.
Another difficulty of the DiagMC approach is that it can be challenging to
resum the perturbation series and obtain converged results even if many
coefficients are known with great accuracy.

In order to reduce the sign problem of the DiagMC a connected determinant
algorithm (CDet) has been recently introduced in Ref.~\onlinecite{rossi}.  The
key idea of the approach is to express the sum of a factorial number of
connected perturbation diagrams as a sum of determinants (a similar strategy is
used in an algorithm for correlated out-of-equilibrium
systems).~\cite{parcollet_prb_2015}  The physical quantity of interest is then
obtained by stochastically sampling these contributions. This algorithm has
been shown to scale as $3^n$ with the perturbation order $n$.  It has proven to
give quantitative improvements in the computation of static properties such as
pressure.~\cite{rossi} However, no computation of dynamical quantities with the
CDet approach has been attempted so far.

In this article, we introduce and compare three different Monte Carlo
determinantal algorithms that allow to compute dynamical quantities of a
fermionic system.  Two of them are directly based on the CDet approach, while
the third algorithm, which we will show is the most efficient, is a
generalization of the CDet approach to one-particle-irreducible (1PI) diagrams.
It directly samples the contributions to the self-energy with a recursive
algorithm scaling as $n^2 3^n$. By comparing results for the two-dimensional
Hubbard model with those obtained from DiagMC, we will show that this new
approach leads to much smaller error bars for the same numerical effort. It
therefore represents an important alternative to compute dynamical quantities.

The article is organized  as follows. In Sec.~\ref{sec:cdet}, we briefly
summarize the CDet approach introduced in Ref.~\onlinecite{rossi} as it will be
one of the building blocks of our proposal.  In Sec.~\ref{sec:dynamical}, we
present three algorithms that allow one to derive dynamical quantities.  We discuss
their practical implementation as a Monte Carlo method in
Sec.~\ref{sec:montecarlo}. We then compare and discuss the results of these
algorithms and of the DiagMC for the two-dimensional Hubbard model in
Sec.~\ref{sec:results}. We finally conclude in Sec.~\ref{sec:conclusion}.


\section{Connected determinant approach}\label{sec:cdet}

First, we briefly summarize the CDet approach introduced in
Ref.~\onlinecite{rossi} as it is one of the building blocks of our proposed
algorithms. This approach provides a general scheme to compute connected
correlators.  For concreteness, we consider in this article models described by
a non-interacting Green's function $G_0$ and a local interaction vertex $U
n_{\uparrow} n_{\downarrow}$. This is the case for example in the Hubbard
model, in some quantum impurity problems or in the simple case of an isolated
Hubbard atom (that we will later use for benchmark purposes).

In a diagrammatic approach, a perturbation series in the interaction $U$ is
constructed. Correlation functions $\mathcal{C}$ of two operators $A$ and $B$, defined as
\begin{equation}
  \mathcal{C}(\xout, \xin) \equiv - \langle T_\tau B(\xout) A(\xin) \rangle,
\end{equation}
where $T_\tau$ is the time-ordering operator and $x$ denotes a vertex, are then expressed as a sum of \emph{connected} diagrams. 
In real space and imaginary time,  $x$ writes $(i,\tau)$ for the Hubbard model, where 
$i$ describes the lattice position and $\tau \in [0, \beta]$ the imaginary time ($\beta = 1/T$ being the inverse temperature).
At a given order $n$ in the perturbation series, a diagram contributing to $\mathcal{C}(\xout, \xin)$ is 
characterized  by the set of its internal vertices $V = \{x_1, ..., x_n\}$ where $x_l$ is associated with the $l$-th interaction vertex.
The topology of such a diagram is given by two adjacency matrices describing the way 
the interaction vertices and the external vertices $\xin$ and $\xout$ are connected.

In the standard DiagMC~\cite{prokofev_prl_1998, prokofev_prl_2007,
burovski_prb_2004, vanhoucke_2010, kozik_epl_2010} technique, individual
connected diagrams are stochastically sampled in a way that preserves their
connectivity, with a probability given by the absolute value of their
contribution to $\mathcal{C}(\xout, \xin)$.
Note that even if some diagrams share the same vertices, they may have alternating signs from
one topology to another, which is one of the ingredients leading to a significant sign
problem in this approach.
The idea of the CDet algorithm  is to regroup \emph{all} diagrams sharing the
same internal vertices $V$ in a contribution $\mathcal{C}_V(\xout, \xin)$, and
then stochastically sample the sets $V$. The stochastic weight of this group
of diagrams in the Monte Carlo sampling of $\mathcal{C}(\xout, \xin)$ is the
absolute value of their sum, which is only a function of $V$.

One could naturally expect that summing this factorial number of diagrams would come with
a factorial cost, but it was shown~\cite{rossi} that it can actually be achieved
exponentially. The sum of connected diagrams entering $\mathcal{C}_V(\xout, \xin)$ is expressed as the
sum of \emph{all} diagrams (connected and disconnected ones) from which the disconnected components are
recursively subtracted . This can be formalized  as follows
\begin{fmffile}{introduction}
\begin{equation}\label{rossi_eq}
\begin{split}
\footnotesize \mathcal{C}_V(x_{\text{out}},x_{\text{in}})  & = \mathcal{D}_V(x_{\text{out}},x_{\text{in}}) - \hspace{-0.1cm} \sum_{S \varsubsetneq V} \mathcal{C}_S(x_{\text{out}},x_{\text{in}}) \mathcal{D}_{V \backslash S}(\varnothing), \\ \\
\begin{gathered}
\begin{fmfgraph*}(10,15)
\fmfset{arrow_len}{3mm}
\fmftop{v1}
\fmfbottom{v2}
\fmf{dbl_plain_arrow,label.dist=0,label=\hspace{-0.7cm} \vspace{1cm}$\mathcal{C}$}{v1,v2}
\fmf{phantom,label.dist=0,label=\hspace{0.6cm} \vspace{-0.8cm}$V$}{v1,v2}
\fmflabel{$x_{\text{in}}$}{v1}
\fmflabel{$x_{\text{out}}$}{v2}
\end{fmfgraph*}
\end{gathered}
\hspace{0.1cm}  & =
\hspace{0.4cm}
\begin{gathered}
\begin{fmfgraph*}(10,15)
\fmfset{arrow_len}{3mm}
\fmftop{i1}
\fmfbottom{o1}
\fmf{fermion}{i1,v1}
\fmf{phantom,label.dist=0,label=\vspace{0.4cm} \small All vert. in $V$,tension=0.25}{v1,v2}
\fmf{phantom,label.dist=0,label=\vspace{-0.4cm} \small (incl. disc.),tension=0.25}{v1,v2}
\fmf{fermion}{v2,o1}
\fmflabel{$x_{\text{in}}$}{i1}
\fmflabel{$x_{\text{out}}$}{o1}
\end{fmfgraph*}
\end{gathered}
\hspace{0.4cm}- \sum_{S \varsubsetneq V} 
\hspace{-0.3cm}
\begin{gathered}
\begin{fmfgraph*}(10,15)
\fmfset{arrow_len}{3mm}
\fmftop{v1}
\fmfbottom{v2}
\fmf{dbl_plain_arrow,label.dist=0,label=\hspace{-0.6cm} \vspace{1cm}$\mathcal{C}$}{v1,v2}
\fmf{phantom,label.dist=0,label=\hspace{0.6cm} \vspace{-1cm}$S$}{v1,v2}
\fmflabel{$x_{\text{in}}$}{v1}
\fmflabel{$x_{\text{out}}$}{v2}
\end{fmfgraph*}
\end{gathered}
\hspace{-0.1cm}
\times
\hspace{0.5cm}
\begin{gathered}
\begin{fmfgraph*}(10,15)
\fmfset{arrow_len}{3mm}
\fmfleft{i1}
\fmfright{o1}
\fmf{phantom,left, tag=1}{i1,o1}
\fmf{phantom,left, tag=2}{o1,i1}
\fmfipath{p[]}
\fmfiset{p1}{vpath1(__i1,__o1)} 
\fmfiset{p2}{vpath2(__o1,__i1)}
\fmfi{fermion}{subpath (length(p1)/4,3length(p1)/4) of p1} 
\fmfi{fermion}{subpath (length(p2)/4,3length(p2)/4) of p2} 
\fmffreeze
\fmf{phantom,label.dist=0,label= \vspace{0.3cm} \small All vert. in $V \backslash S$}{i1,o1}
\fmf{phantom,label.dist=0,label= \vspace{-0.3cm} \small (incl. disc.)}{i1,o1}
\end{fmfgraph*}
\end{gathered}
\end{split}
\end{equation}
\end{fmffile}
where $\mathcal{D}_V(\xout,\xin)$ denotes the sum of all diagrams (including
disconnected ones) with internal vertices $V$, external vertices $\xin$ and $\xout$. $D_V(\varnothing)$
is the sum of all diagrams with vertices $V$ and no external vertices.
The cancellation of disconnected diagrams is illustrated in the
second line of Eq.~\eqref{rossi_eq}. A key feature of this recursive sum is that $\mathcal{D}_V$ terms  can be
expressed as determinants (and hence with a polynomial computational cost).~\cite{rossi_epl_2017}

Algorithmically, the evaluation of $\mathcal{C}_V(\xout,\xin)$ at order $n$ is
done in two steps. First, determinants $\mathcal{D}_S$ are computed for all
subsets $S$ of $V$, with a total effort $2^n n^3$. The leading complexity
however comes from the progressive computation, from low to high orders, of the
$\mathcal{C}_S$. More precisely, if all $\mathcal{C}_{S'}$ are known for
subsets $S'$ with less than $p \leq n$ vertices, one can compute a given $p$-order
$\mathcal{C}_S$ using Eq.~\eqref{rossi_eq} with $V=S$, in $2^p$ operations (see r.h.s of
Eq.~\eqref{rossi_eq}). This has to be done for all the $\binom{n}{p}$ subsets
$S$ at order $p$ before computing contributions at the next order $p+1$. The
final result is obtained when this has been done for all $p \le n$ and the
leading complexity of the algorithm to compute $\mathcal{C}_V(\xout,\xin)$ is
therefore $\sum_{p=0}^n \binom{n}{p} 2^p = 3^n$.

Note that a similar cancellation of disconnected diagrams had been introduced
in a quantum Monte Carlo algorithm for correlated out-of-equilibrium
systems~\cite{parcollet_prb_2015} where connected correlators are expressed as
a sum of $2^n$ determinants thanks to Keldysh diagrammatic
techniques.

The CDet approach leads to an important reduction of statistical error with
respect to the DiagMC and has allowed for great progress in the computation of
static properties, such as pressure in the Hubbard model.~\cite{rossi} This
method, however, has not yet been used to compute dynamical quantities.

In the following, we will investigate how this can be done.  We could examine,
e.g. the Green's function but choose instead to focus on the self-energy
$\Sigma$ that is a more irreducible object where signatures of numerical noise
are clearer. Other single-particle quantities can then be computed from
$\Sigma$. A straightforward way to obtain $\Sigma$ is to compute the Green's
function $G$ with the CDet approach and derive the self-energy through Dyson's
equation. However, even if one can compute $G$ with great accuracy, its
inversion in Dyson's equation leads to an amplification of the statistical
noise and, as we will show below, the resulting $\Sigma$ can only be accurately
obtained for low orders.  It is therefore desirable to look for other
techniques to compute the self-energy.  This is the purpose of the following
section.


\section{Self-energy computation}\label{sec:dynamical}

We introduce three different techniques to compute dynamical quantities.  In
order to compare their efficiencies, we focus on the self-energy
$\Sigma^\sigma$ (here $\sigma$ denotes the spin) that they yield, because
numerical noise is particularly visible in this quantity. First, we use Dyson's
equation to obtain the self-energy from a computation of the Green's function
using the CDet technique. We then present a diagrammatic method that allows us to
compute the self-energy recursively from the knowledge of a different
correlator $\bar{F}$ that can still be computed using the CDet. We finally
introduce an extension of the CDet algorithm that efficiently computes the sum
of all \emph{one-particle-irreducible} diagrams of a perturbation series and
therefore allows us to directly stochastically sample the contributions to the
self-energy.  As we discuss in Sec.~\ref{sec:results}, the latter allows
for a much better determination of dynamical quantities.

\subsection{Dyson's equation}

The most straightforward way to compute the self-energy $\Sigma^\sigma$ is to
first compute the Green's function $G^\sigma$ using the CDet algorithm and then
use Dyson's equation
\begin{equation}
  \Sigma^\sigma = (G_0^\sigma)^{-1} - (G^\sigma)^{-1}.
\end{equation}
We show in Sec.~\ref{sec:results} that it is very difficult to obtain precise data
with this method because of the inversion of $G$ that dramatically increases
the noise.

\subsection{Equations of motion}

We present a diagrammatic approach to compute the self-energy based on the
computation of a different correlator with the CDet algorithm.
Let us first write the self-energy as the sum of a constant Hartree term and a
frequency-dependent part
\begin{equation}\label{def_sigma}
  \Sigma^\sigma(\xout,\xin) \equiv \Sigma^{H,\sigma} \, \delta_{\xin,\xout} + \tilde\Sigma^\sigma(\xout,\xin).
\end{equation}
We recall that $x$ is a combined index, e.g. $(i,\tau)$ for the Hubbard
model, where $i$ is the lattice site and $\tau$ the imaginary time. The
Hartree term contribution is given by
\begin{equation}\label{hartree_term}
\begin{fmffile}{hartree_term}
\Sigma^{H,\sigma} \equiv U G^{\bar{\sigma}}(0^-)
= 
\hspace{.4cm}
\begin{gathered}
\begin{fmfgraph*}(8,10)
\fmftop{i}
\fmfbottom{o}
\fmf{phantom}{i,v,o}
\fmf{photon}{o,v}
\fmf{dbl_plain,right,tension=1}{v,i,v}
\fmf{phantom, label.dist=0,label=\hspace{1.7cm} \vspace{.5cm}$G^{\bar{\sigma}}(0^-)$,tension=.2}{i,o}
\fmfdot{o}
\end{fmfgraph*}
\end{gathered}
\end{fmffile}
\end{equation}
It can be directly computed from the knowledge of the Green's function
$G^{\bar{\sigma}}$ which is a connected correlator that can be obtained from Eq.~\eqref{rossi_eq}.
The self-energy $\Sigma^\sigma$ can then be obtained recursively using the
following expression
\begin{subequations}\label{eq:sigmafbar}
\label{eom}
\begin{fmffile}{eom}
\begin{align}
\Sigma^\sigma \hspace{0.1cm}& =\Sigma^{H,\sigma} + \hspace{0.3cm} \bar{F}^\sigma
 - \Sigma^\sigma G^\sigma \Sigma^\sigma, \\
\begin{gathered}
\begin{fmfgraph*}(10,10)
\fmftop{v1}
\fmfbottom{v2}
\fmf{plain,right,tension=1}{v1,v2,v1}
\fmf{phantom, label.dist=0,label=$\Sigma^\sigma$}{v1,v2}
\end{fmfgraph*}
\end{gathered}\hspace{0.2cm}& =
\begin{gathered}
\begin{fmfgraph*}(8,10)
\fmftop{i}
\fmfbottom{o}
\fmf{phantom}{i,v,o}
\fmf{photon}{o,v}
\fmf{dbl_plain,right,tension=1}{v,i,v}
\fmf{phantom, label.dist=0,label=\hspace{1cm} \vspace{.5cm}$G^{\bar{\sigma}}$,tension=.2}{i,o}
\fmfdot{o}
\end{fmfgraph*}
\end{gathered} 
\hspace{0.2cm}
+
\hspace{0.6cm}
\begin{gathered}
\begin{fmfgraph*}(10,22)
       \fmftop{i1,i2}
       \fmfbottom{o1,o2}
	\fmf{photon}{i1,i2}
	\fmf{photon}{o1,o2}
	\fmf{fermion}{i1,v1}
	\fmf{phantom, label.dist=0,label= \hspace{1.1cm} \vspace{0.4cm} All vertices in $V$, tension=0.26}{v1,v2}
	\fmf{phantom, label.dist=0,label= \hspace{1.1cm} \vspace{-0.35cm} (connected), tension=0.26}{v1,v2}
	\fmf{fermion}{v2,o1}
	\fmfdot{i1}
	\fmfdot{o1}
	\fmf{phantom,left=0.5,tension=0.2, tag=1}{i2,o2}
	\fmf{phantom,left=0.5,tension=0.2, tag=2}{o2,i2}
	\fmfipath{p[]}
	\fmfiset{p1}{vpath1(__i2,__o2)} 
	\fmfiset{p2}{vpath2(__o2,__i2)}
	\fmfi{fermion}{subpath (0,length(p1)/4) of p1} 
	\fmfi{fermion}{subpath (3length(p1)/4,length(p1)) of p1} 
	\fmfi{fermion}{subpath (0,length(p2)/4) of p2} 
	\fmfi{fermion}{subpath (3length(p2)/4,length(p2)) of p2}
     \end{fmfgraph*}
\end{gathered}
\hspace{0.8cm}
 - 
 \hspace{-0.2cm}
 \begin{gathered}
  \begin{fmfgraph*}(10,20)
       \fmftop{tin}
       \fmfbottom{tout}
       \fmf{plain,right,tension=.3}{tin,t1,tin}
       \fmf{phantom, label.dist=0,label=$\Sigma^\sigma$}{tin,t1}
       \fmf{dbl_plain, label.dist=0, tension=2}{t1,t3}
       \fmf{phantom, label.dist=0, label=\hspace{0.6cm} \vspace{0.5cm} $G^\sigma$}{t2,t3}
        \fmf{plain,right,tension=.3}{t3,tout,t3}
       \fmf{phantom, label.dist=0,label=$\Sigma^\sigma$}{t3,tout}
     \end{fmfgraph*}
     \end{gathered}
\end{align}
\end{fmffile}
\end{subequations}
where the correlation function $\bar{F}^\sigma$ is defined by
\begin{equation}\label{def_fbar}
  \bar{F}^\sigma(\xout,\xin) \equiv -U^2 \langle T_\tau n_{\bar{\sigma}}c_\sigma(\xout) n_{\bar{\sigma}}c_\sigma^\dagger(\xin) \rangle.
\end{equation}
Equation~\eqref{eq:sigmafbar} can be derived from the equations of motion (EOM) of
the Green's function, as detailed in Appendix~\ref{app:eom}, and we will use
this terminology in the following to unambiguously refer to this method. It has
a simple diagrammatic interpretation, see the second line of Eq.~\eqref{eom},
that illustrates how 1PI diagrams are isolated. Indeed, according to Eq.
\eqref{def_sigma}, the self-energy is the sum of contributions with a single
external vertex (Hartree term $\Sigma^{H,\sigma}$) and contributions with two
external vertices ($\tilde{\Sigma}^\sigma$). The former is easy to compute, and
the latter is the sum of all 1PI diagrams with two external vertices. The term
$\bar{F}^\sigma$ on the r.h.s of Eq.~\eqref{eom} represents the sum of
\emph{all connected} diagrams with the same external vertices as
$\tilde{\Sigma}^\sigma$. From this, one then has to subtract all non-1PI
diagrams, which can always be expressed in the form $\Sigma^\sigma G^\sigma
\Sigma^\sigma$.

We now reorganize the equation above in order to be able to compute the
contributions to the self-energy at a given perturbation order just from the
knowledge of the contributions to $\bar{F}^\sigma$ and $\Sigma^{H,\sigma}$. We
first multiply Eq.~\eqref{eq:sigmafbar} by $G_0^\sigma$ on the right and we
obtain
\begin{equation}
\Sigma^\sigma G_0^\sigma = \Sigma^{H,\sigma} G_0^\sigma + \bar{F}^\sigma G_0^\sigma - \Sigma^\sigma G^\sigma \Sigma^\sigma G_0^\sigma.
\end{equation}
Reorganizing the terms,
\begin{equation}
\bar{F}^\sigma G_0^\sigma  + \Sigma^{H,\sigma} G_0^\sigma = \Sigma^\sigma[G_0^\sigma+G^\sigma \Sigma^\sigma G_0^\sigma] = \Sigma^\sigma G^\sigma .
\end{equation}
Substituting this expression for $\Sigma^\sigma G^\sigma$ in Eq.~\eqref{eom}, we find
\begin{equation}\label{eom_final}
\Sigma^\sigma = \Sigma^{H,\sigma} + \bar{F}^\sigma - [\bar{F}^\sigma G_0^\sigma + \Sigma^{H,\sigma} G_0^\sigma]\Sigma^\sigma.
\end{equation}
This equation allows us to recursively compute the contributions to the
self-energy at all perturbation orders. Indeed, because $\bar{F}^\sigma$ is at
least of order 2 in $U$ and $\Sigma^{H,\sigma}$ is at least of order 1 in $U$,
the computation of the contribution to the self-energy at order $n$ on the
l.h.s can be obtained from the knowledge of the contributions to $\bar{F}$ and
the contributions to the self-energy at strictly lower orders $< n$ on the
r.h.s. As a result, the l.h.s contributions can be computed without any
inversion and there is no noise amplification as in Dyson's equation.  We can
therefore expect this approach to be more efficient.

The algorithm is implemented by computing the Green's function $G^\sigma$ and
the correlator $\bar{F}^\sigma$ using the CDet algorithm.  Then,
Eq~\eqref{eom_final} is used to recursively compute the contributions to
$\Sigma^\sigma$ at a given order.  As we use the CDet algorithm to obtain two
correlators, and the self-energy is only computed in a post-processing part,
the complexity of this algorithm naturally scales as $3^n$.

\subsection{Determinantal approach to sum all 1PI diagrams}

We now introduce an extension of the CDet algorithm to efficiently compute the sum of
all one-particle irreducible diagrams of a perturbation series.
At a given perturbation order $n$ in the interaction $U$, a self-energy diagram
is characterized  by $\xin$, $\xout$, its internal interaction vertices $V =
\{x_1, \ldots, x_{n-2}\}$, and the adjacency matrices that connect the
vertices.  Note that we choose $n-2$ points in the set of internal vertices $V$
because $\xin$ and $\xout$ both carry an interaction vertex as well.
We wish to group all diagrams that share the same internal vertices
$V$ into a contribution $\Sigma_V^\sigma(\xout, \xin)$ so that
\begin{align}
\label{eq:sigma_sum_V}
  \Sigma^\sigma(\xout, \xin) &= \sum_V \Sigma_V^\sigma(\xout, \xin) \\ \nonumber
   &= \sum_V \left( \Sigma_V^{H,\sigma} \, \delta_{\xin,\xout} + \tilde\Sigma_V^\sigma(\xout,\xin) \right).
\end{align}
The contribution $\Sigma_V^\sigma(\xout, \xin)$ is theoretically a sum of a factorial
number of diagrams, but we will express it with the help
of a recursion, very much in the
spirit of Ref.~\onlinecite{rossi}, that only involves connected correlators that can be computed with
exponential effort using Eq.~\eqref{rossi_eq}. The numerical effort to
obtain $\Sigma_V^\sigma(\xout, \xin)$ will then turn out to also be exponential.

The frequency-dependent part of the self-energy $\tilde\Sigma_V^\sigma(\xout,\xin)$
can be expressed \emph{via} the following recursive formula
\begin{widetext}
\begin{subequations}
\label{direct_self}
\begin{fmffile}{main_equation}
\begin{align}
\tilde\Sigma_V^\sigma(\xout, \xin) \hspace{0.1cm}& = \hspace{0.3cm} \bar{F}_V^\sigma(\xout, \xin) \hspace{0.5cm} - \hspace{-0.2cm}
\sum_{\substack{x' \in V \\ S \subseteq V \backslash \{x'\} \\ S' = V \backslash \left(S \cup \{ x'\}\right)}} \hspace{-0.6cm} F_{S'}^\sigma(\xout, x')\tilde\Sigma_S^\sigma(x',\xin) \hspace{0.2cm} -  \sum_{\substack{S \subseteq V \\  S' = V \backslash S}}F_{S'}^\sigma(\xout, \xin) \left(UG_S^{\bar{\sigma}}(0^-)\right), \\
\begin{gathered}
\begin{fmfgraph*}(10,10)
\fmftop{v1}
\fmfbottom{v2}
\fmf{plain,right,tension=1,fore=red}{v1,v2,v1}
\fmf{phantom, label.dist=0,label=$\tilde\Sigma_V^\sigma$}{v1,v2}
\fmfdot{v1,v2}
\fmflabel{$\xin$}{v1}
\fmflabel{$\xout$}{v2}
\end{fmfgraph*}
\end{gathered}\hspace{0.2cm}& =
\hspace{0.6cm}
\begin{gathered}
\begin{fmfgraph*}(10,22)
       \fmftop{i1,i2}
       \fmfbottom{o1,o2}
	\fmf{photon}{i1,i2}
	\fmf{photon}{o1,o2}
	\fmf{fermion}{i1,v1}
	\fmf{phantom, label.dist=0,label= \hspace{1.1cm} \vspace{0.4cm} All vertices in $V$, tension=0.26}{v1,v2}
	\fmf{phantom, label.dist=0,label= \hspace{1.1cm} \vspace{-0.35cm} (connected), tension=0.26}{v1,v2}
	\fmf{fermion}{v2,o1}
	\fmfdot{i1}
	\fmfdot{o1}
	\fmflabel{$x_{\text{in}}$}{i1}
        \fmflabel{$x_{\text{out}}$}{o1}
	\fmf{phantom,left=0.5,tension=0.2, tag=1}{i2,o2}
	\fmf{phantom,left=0.5,tension=0.2, tag=2}{o2,i2}
	\fmfipath{p[]}
	\fmfiset{p1}{vpath1(__i2,__o2)} 
	\fmfiset{p2}{vpath2(__o2,__i2)}
	\fmfi{fermion}{subpath (0,length(p1)/4) of p1} 
	\fmfi{fermion}{subpath (3length(p1)/4,length(p1)) of p1} 
	\fmfi{fermion}{subpath (0,length(p2)/4) of p2} 
	\fmfi{fermion}{subpath (3length(p2)/4,length(p2)) of p2}
     \end{fmfgraph*}
\end{gathered}
\hspace{1cm}
 - \hspace{-0.2cm}\sum_{\substack{x' \in V \\ S \subseteq V \backslash \{x'\} \\ S' = V \backslash \left(S \cup \{ x'\}\right)}} 
\hspace{0.4cm}
 \begin{gathered}
  \begin{fmfgraph*}(10,20)
       \fmftop{tin}
       \fmfbottom{tout}
       \fmf{plain,right,tension=.3,fore=red}{tin,t1,tin}
       \fmf{phantom, label.dist=0,label=$\tilde\Sigma_S^\sigma$}{tin,t1}
       \fmfdot{tin}
       \fmfdot{tout}
       \fmfdot{t1}
       \fmflabel{$x_{\text{in}}$}{tin}
       \fmflabel{$x_{\text{out}}$}{tout}
       \fmf{dbl_plain, label.dist=0, label= \hspace{-0.5cm} \footnotesize $x'$, tension=5, fore=blue}{t1,t2}
       \fmf{dbl_plain, label.dist=0, label=\hspace{1.8cm} \vspace{-0.4cm} \Huge \} \hspace{-0.4cm} \normalsize $F_{S'}^\sigma$, tension=5, fore=blue}{t2,t3}
       \fmf{phantom, label.dist=0, label=\hspace{0.5cm} \vspace{0.2cm} $G^\sigma$}{t2,t3}
        \fmf{plain,right,tension=.5, fore=blue}{t3,tout,t3}
       \fmf{phantom, label.dist=0,label=$\Sigma^\sigma$}{t3,tout}
     \end{fmfgraph*}
     \end{gathered}
\hspace{1.5cm}- \sum_{\substack{S \subseteq V \\ S' = V \backslash S}}
\hspace{1.1cm}
\begin{gathered}
\begin{fmfgraph*}(10,15)
	\fmftop{i1,i2}
	\fmfbottom{o1,o2}
        \fmf{phantom}{i1,v1,i2}
        \fmf{phantom}{o1,v2,o2}
        \fmf{phantom,tension=0}{v1,v2}
	\fmf{photon}{i1,v1}
	\fmf{dbl_plain,right,tension=1}{v1,i2,v1}
	\fmf{phantom, label.dist=0,label=\hspace{2cm}$G_S^{\bar{\sigma}}(0^-)$,tension=.2}{v1,i2}
	\fmf{dbl_plain,label.dist=0,label=\hspace{-0.6cm}\vspace{-0.3cm}$G^\sigma$, fore=blue}{i1,v3}
	\fmf{dbl_plain, label.dist=0,label=\hspace{1.8cm} \vspace{-0.4cm} \Huge \} \normalsize $F_{S'}^\sigma$, fore=blue}{v3,v4}
	\fmf{plain,right,tension=.1, fore=blue}{v4,o1,v4}
	\fmf{phantom, label.dist=0,label=$\Sigma^\sigma$,tension=.4}{v4,o1}
	\fmfdot{i1}
	\fmfdot{o1}
	\fmflabel{$x_{\text{in}}$}{i1}
        \fmflabel{$x_{\text{out}}$}{o1}
    \end{fmfgraph*}
\end{gathered}
\end{align}
\end{fmffile}
\end{subequations}
\end{widetext}
where the correlation function $F^\sigma$ is given by~\cite{bulla_1998}
\begin{align}\label{def_f}
  F^\sigma(\xout,\xin) &= \Sigma^\sigma G^\sigma (x_{\text{out}}, x_{\text{in}}) \\
                       &\equiv -U\langle T_\tau n_{\bar{\sigma}}c_\sigma(x_{\text{out}}) c_\sigma^\dagger(x_{\text{in}})\rangle,
\end{align}
and $\bar{F}^\sigma$ by Eq.~\eqref{def_fbar}.
The starting point of the recursion is the order-2 diagram
\begin{equation}
\label{start_recursion}
\begin{fmffile}{start_recursion}
\begin{gathered}
\begin{fmfgraph*}(8,8)
\fmfleft{i}
\fmfright{o}
\fmf{plain,right,tension=1}{i,o,i}
\fmf{phantom, label.dist=0,label=$\tilde\Sigma^\sigma_\varnothing$}{i,o}
\fmfdot{i,o}
\fmflabel{$x_{\text{in}}$}{i}
\fmflabel{$x_{\text{out}}$}{o}
\end{fmfgraph*}
\end{gathered}
\hspace{.8cm}
= 
\hspace{.4cm}
\begin{gathered}
\begin{fmfgraph*}(13,8)
\fmfleft{i1,i2}
\fmfright{o1,o2}
\fmf{photon}{i1,i2}
\fmf{photon}{o1,o2}
\fmf{fermion}{i1,o1}
\fmf{fermion,left=0.5,tension=0.2, tag=1}{i2,o2,i2}
\fmfdot{i1}
\fmfdot{o1}
\fmflabel{$x_{\text{in}}$}{i1}
\fmflabel{$x_{\text{out}}$}{o1}
\end{fmfgraph*}
\end{gathered}
\end{fmffile}
\end{equation}

The second line of Eq.~\eqref{direct_self} illustrates the cancellation of non-1PI diagrams.
The self-energy contributions $\tilde\Sigma_V^\sigma$ that
are calculated recursively are indicated as red circles, while blue diagrams
correspond to the correlation function $F^\sigma=\Sigma^\sigma G^\sigma$. An explicit example of this formula at third order is shown in
Appendix~\ref{explicit}. Let us note that, in this
 formula, the starting point of the recursion is already an order-2 diagram
 while it is an order-0 diagram in Eq.~\eqref{rossi_eq}, justifying a set $V$ with $n-2$ vertices.

The first term $\bar{F}_V^\sigma(\xout, \xin)$
on the r.h.s of Eq.~\eqref{direct_self} is the contribution
to the correlation function $\bar{F}^\sigma(\xout,\xin)$
for the set of internal vertices $V$. It is the sum of
all connected diagrams that have interaction vertices at $\xin$, $\xout$ and
all $x \in V$ as interaction vertices. In order to obtain the contributions to the self-energy
$\tilde\Sigma^\sigma_V(\xout,\xin)$, one has to subtract from this term all diagrams that
are not 1PI. These ones can be
expressed in the form $\Sigma^\sigma G^\sigma \Sigma^\sigma$ = $F^\sigma \Sigma^\sigma$ = $F^\sigma (\Sigma^{H,\sigma}
+ \tilde\Sigma^\sigma)$ and there are therefore
two families of diagrams to subtract for a given set of vertices $V$: first all terms
$F^\sigma_{S'}(\xout,\xin) \Sigma^{H,\sigma}_S$ such
that $S \sqcup S' = V$, then all terms
$F^\sigma_{S'}(\xout,x') \tilde\Sigma^\sigma_S(x',\xin)$ such
that $S \sqcup \{x'\} \sqcup S' = V$. In the latter family, note that $S \varsubsetneq V$
is a proper subset of $V$, so that the calculation of
$\tilde\Sigma^\sigma_V$ involves only some $\tilde\Sigma^\sigma_S$
that have been previously computed in the recursion.

We have therefore derived a recursive formula for the contributions
$\tilde{\Sigma}^\sigma_V(\xout,\xin)$ that involves the computation of only connected
correlation functions. The recursion is completed in two steps. First, all
correlators $\bar{F}^\sigma$, $F^\sigma$ and $G^\sigma$ have to be enumerated, the main effort coming
from the $F^\sigma_S$ that have to be computed for all pairs of
external vertices (as a consequence of the explicit use of an intermediate
vertex point $x'$ in Eq.~\eqref{direct_self}). The computational cost for the
precomputation is therefore dominated by $n^2 3^n$.
Second, the recursion has to be implemented, as in the CDet, by computing the
contributions $\tilde\Sigma_S^\sigma$ starting from low to higher orders. At a given
order $p$, it takes an effort $p 2^p$ to get a given
$\tilde\Sigma_S^\sigma(x',\xin)$.  This has to be done for all subsets $S$ at order
$p$ and all $x'$ before computing contributions at the next order $p+1$ and
requires a total effort $\binom{n}{p} p^2 2^p$. All in all the recursion will
take $\sum_{p=0}^n \binom{n}{p} p^2 2^p$ with a complexity $n^2 3^n$.
The leading complexity of the algorithm is therefore $n^2 3^n$.

We will show in Sec.~\ref{sec:results} that despite this additional $n^2$ factor,
this method leads to smaller error bars compared to the approaches above. It
also gives more accurate results than the state-of-the-art DiagMC calculations
for the same computational effort.

\section{Monte Carlo implementation}\label{sec:montecarlo}

In this section, we describe how to compute the different quantities that
appear in the algorithms above using a Monte Carlo (MC) method.
We generically denote these quantities as $\mathcal{M}^\sigma$. The quantities that need
to be computed depend on the algorithm considered. The Green's function
$G^\sigma$ has to be computed for all three approaches. In addition
$\bar{F}^\sigma$ must be computed for the equations of motion algorithm and
$\tilde\Sigma^\sigma$ for the direct sampling of the self-energy.  We write
$\mathcal{M}^\sigma$ as a sum over all contributions described by a set $V_m$
with $m$ internal vertices
\begin{equation}
  \mathcal{M}^\sigma(\xout, \xin) = \sum_{m=0}^\infty \sum_{V_m} \mathcal{M}^\sigma_{V_m}(\xout,\xin).
\end{equation}
Note that a configuration with $m$ internal vertices contributes, in the
perturbation series in $U$, to the coefficient of order $n = m$ for the Green's
function, $n = m+1$ for $F^\sigma$ and $n = m+2$ for $\tilde\Sigma^\sigma$.

In order to compute $\mathcal{M}^\sigma(\xout, \xin)$, we
stochastically generate Monte Carlo configurations that sample the r.h.s terms of the sum.
A configuration $\mathfrak{C}$ is described by the number of internal vertices
$m$, the spin $\sigma$ and the set of all vertices
\begin{equation}
  \mathfrak{C} = \{m; \sigma; \xin, \xout; x_1, \dots, x_m \},
\end{equation}
and its weight in the Monte Carlo sampling is
\begin{equation}
  \text{\emph{w}}_{\mathfrak{c}} = \left| \mathcal{M}^\sigma_{V_m}(\xout, \xin) \right|.
\end{equation}

We use a standard Metropolis~\cite{metropolis_1953} algorithm to generate a Markov
chain distributed according to $\text{\emph{w}}_{\mathfrak{c}}$. For concreteness, we consider
the case of the Hubbard model where $x = (i,\tau)$. Starting from
a given $\mathfrak{C}$, a new configuration $\mathfrak{C}'$ is proposed by applying one
of the following Monte Carlo updates:
\begin{enumerate}

  \item Pick one of the interaction vertices in $\mathfrak{C}$ and change its
        position and imaginary time. One can increase the probability of the move being
        accepted by choosing  a new position either among the neighbors  of the chosen
        vertex or from a Gaussian distribution. The imaginary time can be chosen
        uniformly.

  \item Flip the spin $\sigma \rightarrow \bar \sigma$.

  \item Remove a randomly chosen internal interaction vertex from $\mathfrak{C}$.

  \item Add a new internal interaction vertex in $\mathfrak{C}$.
        The new lattice site can be chosen from a Gaussian distribution around
        the center of gravity of the vertices in $\mathfrak{C}$.
        The imaginary time can be chosen with uniform probability. 

\end{enumerate}
The new configuration $\mathfrak{C}'$ is accepted or rejected with the usual Metropolis
ratio
\begin{equation}
  p^\mathrm{accept}_{\cc \rightarrow \cc'} =
  \min \left( 1, \frac{T_{\cc' \cc} \, \wcp}{T_{\cc \cc'} \, \wc} \right),
\end{equation}
where $T_{\cc \cc'}$ is the probability to propose $\cc'$ after $\cc$.

This algorithm will sample the configurations according to the weights
$\wc$, however it is necessary to normalize  the result.  To do so, it is
convenient to restrict the Monte Carlo simulation to only two consecutive
orders, $m$ and $m+1$. A vertex can be added (resp. removed) only if the
current $\cc$ is at order $m$ (resp. $m+1$). In the lowest order $m$ the following normalization 
quantity is measured
\begin{equation}
  \mathcal{N}_m = \sum_{\xin, \xout, \sigma} \sum_{V_m}
                  \left| \mathcal{M}^\sigma_{V_m} (\xout, \xin) \right|,
\end{equation}
while at order $m+1$, both $\mathcal{N}_{m+1}$ and the contribution to
$\mathcal{M}^\sigma$ are measured.
The knowledge of the expected value for $\mathcal{N}_m$ allows us to find the
normalization  factor and obtain a normalized  value for the contribution
to $\mathcal{M}^\sigma$ and
$\mathcal{N}_{m+1}$ at order $m+1$.  The latter can then be used to normalize  a
further simulation at orders $m+1$ and $m+2$, and so on. The contribution at $m=0$, for instance the
pair-bubble diagram for the self-energy, can be computed analytically, allowing for a
precise determination of $\mathcal{N}_0$.

We performed several calculations for the special case of a single
correlated site (especially for benchmarking purposes). In that situation, it
is possible to restrict the simulation to a fixed order $m$ and propose updates
that only change the spin $\sigma$ and the imaginary time of a randomly chosen
interaction vertex. The normalization  is obtained by computing an integral
whose value is known. The simple choice
\begin{equation*}
  \mathcal{I}_m = \sum_\sigma
  \int_0^\beta \mathrm{d}\tau_\mathrm{in} \mathrm{d}\tau_\mathrm{out}
       \mathrm{d}\tau_1 \dots \mathrm{d}\tau_{m} = 2 \beta^{m+2}
\end{equation*}
turns out to provide a good normalization .

Let us note that statistical errors in the normalization factor
propagate from one order to the other. One must therefore be careful in the
computation of error bars using, e.g. a binning or jackknife analysis.

\section{Results}\label{sec:results}

In this section, we present actual computations of the self-energy according to
the implementations described in Sec.~\ref{sec:dynamical}. 
For clarity, we respectively denote by Dyson, EOM and $\Sigma$Det the use of Dyson's equation, of the equations of motion,
and of the direct calculation of the self-energy from the sum of 1PI diagrams.

We consider two models in the following. The first is a single correlated
electronic level, that we will refer to as a Hubbard atom, described by the Hamiltonian
\begin{equation}
  \mathcal{H}_\mathrm{atom} = U n_\uparrow n_\downarrow + \epsilon,
\end{equation}
where $n_\sigma$ is the number of the spin-$\sigma$ fermion, $U$ is the onsite
repulsion  and $\epsilon$ the energy of the electronic level.
This model has an analytical solution
and allows us to both benchmark and compare the different methods introduced above.
The second model is the prototypical two-dimensional Hubbard model given by
\begin{equation}
  \mathcal{H}_\mathrm{Hubbard} = -t\sum_{\langle i,j \rangle \sigma} c_{i\sigma}^\dagger c_{j\sigma}
    + U\sum_i n_{i\uparrow}n_{i\downarrow},
\end{equation}
where $c_{i\sigma}^\dagger$ creates a spin-$\sigma$ electron on the site $i$ of
a square lattice, $t > 0$ is the nearest-neighbor hopping and $U$ is the onsite
interaction.  This is the model that we eventually aim to solve in its
thermodynamic limit (infinite lattice).  In our results, $t=1$ will be our
energy unit.  Note that in the computations of the Hubbard model, we use an
$\alpha$ shift that redefines the non-interacting
propagator.~\cite{rubtsov_prb_2005,parcollet_prb_2015,rossi_prb_2016,wei_prb_2017}

We first benchmark our results against both analytical and standard
DiagMC~\cite{prokofev_prl_1998, prokofev_prl_2007, kozik_epl_2010,
vanhoucke_2010, burovski_prb_2004} solutions and verify the theoretical
complexity of our models in Appendix~\ref{sec:bench}. We then compare the three
different methods between them, showing that $\Sigma$Det performs better both on
the isolated atom and on the lattice.  This method is finally shown to also
improve state-of-the-art results from recent DiagMC calculations.

\begin{figure}
\centering
\includegraphics[width=0.47\textwidth]{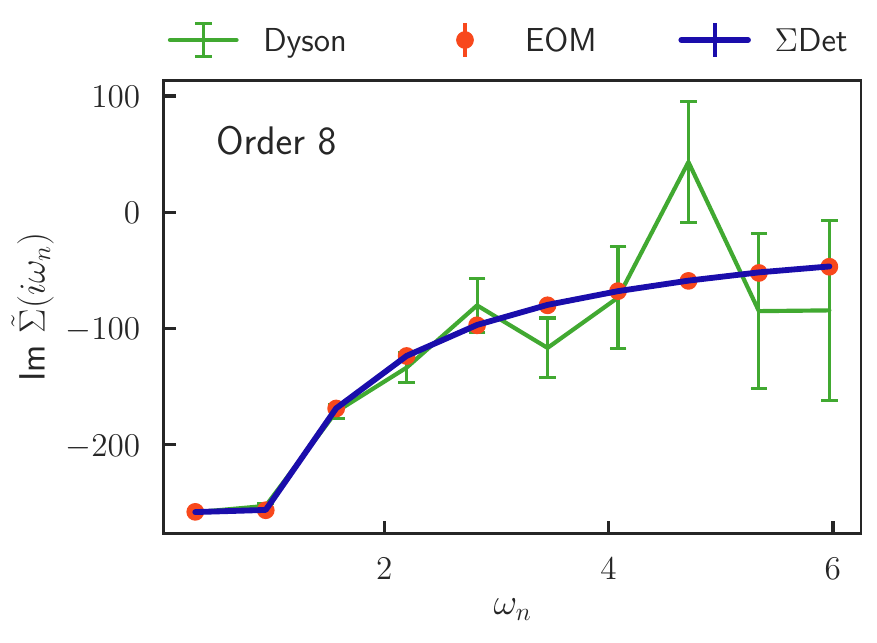}
\vspace{-10pt}
\caption{Imaginary part of the Hubbard atom self-energy
at order 8 in $U$ as obtained from Dyson's equation (green),
the equations of motion approach (orange) and the direct self-energy
measurement (blue). We use $\beta = 10$, $U = 1$, $\epsilon=-0.2$. All
simulations lasted 120 CPU hours.}
\label{comp_dyson_atomic}
\end{figure}

\subsection{Comparison with Dyson's equation}\label{comp_dyson}

Until now, no dynamical quantities have been
computed with the CDet algorithm and it is therefore instructive to see how
the use of Dyson's equation compares to the calculation of the self-energy from the 
EOM and $\Sigma$Det methods.

We first consider the Hubbard atom. Figure~\ref{comp_dyson_atomic} shows the
contribution to the imaginary part of the Matsubara frequency self-energy
$\tilde{\Sigma}^\sigma(i\omega_n)$ from perturbation order 8. The direct
measurement of the self-energy and the EOM method yield results that have very
small error bars (smaller than the symbol size) and that are in perfect
agreement (both curves lie on top of one another). In contrast, starting from
the Green's function as obtained by Eq.~\eqref{rossi_eq}, the results for the
self-energy display large statistical errors that increase with the Matsubara
frequency index.  The reason is simple and expected: when Dyson's equation is
used to compute the self-energy, there is an amplification of the numerical
noise because of the inversion of the Green's function. In practice, it becomes
quickly impossible to obtain accurate data. This is problematic, because large
error bars make it very difficult, e.g. to analytically continue the results to
the real axis.

\begin{figure}
\centering
\includegraphics[width=0.5\textwidth]{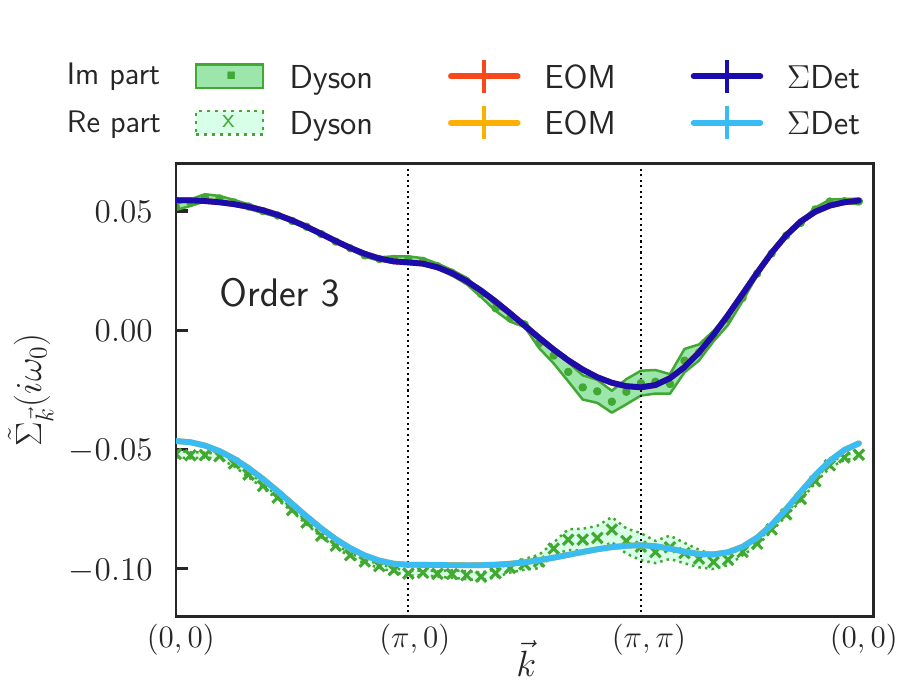}
\vspace{-20pt}
\caption{
Hubbard model self-energy at the first Matsubara frequency
$\tilde{\Sigma}_{\vec{k}}(i\omega_0)$ along the $\vec{k} = (0,0) \rightarrow
(\pi,0) \rightarrow (\pi,\pi) \rightarrow (0,0)$ path at order 3 in $U$, as
obtained from Dyson's equation (green), the equations of motion approach
(orange) and the direct self-energy measurement (blue). We use a $32 \times 32$
lattice with $\beta t = 2$, $U = 4t$, $\mu=0$ and a uniform $\alpha$ shift
$\alpha_{\uparrow} = \alpha_{\downarrow} = 1.53 t$. All simulations lasted 120
CPU hours.}
\label{comp_dyson_lattice}
\end{figure}

Figure~\ref{comp_dyson_lattice} shows results for the two-dimensional Hubbard
model on a $32 \times 32$ lattice (for $\beta t = 2$ the Hubbard model is in
its thermodynamic limit on this lattice). At order 3, the contribution to the
self-energy taken at the first Matsubara frequency $i\omega_0$ obtained from
$\Sigma$Det on a chosen path in the Brillouin zone is in perfect agreement with
the EOM method, and error bars for both methods are very small (smaller than
symbol size, both curves being on top). The computation of $\Sigma^\sigma$ from
the Green's function is noisier. Error bars actually increase with the
Matsubara frequency index when using Dyson's equation, resulting in reasonable
results only for the first few frequencies even for small perturbation orders.
Again, the reason for this large noise is the amplification due to the
inversion of the Green's function. Also, on the lattice, a direct measurement
of the self-energy has the advantage of mainly sampling fairly local diagrams.
Indeed, at a temperature $T = t/2$, the self-energy very quickly vanishes for
non-local components. The same is not true for the Green's function that has a
slower decay; its stochastic sampling is therefore less efficient.

\subsection{Comparison between the equations of motion and the direct sampling of the self-energy}

\begin{figure}
\centering
\includegraphics[width=0.47\textwidth]{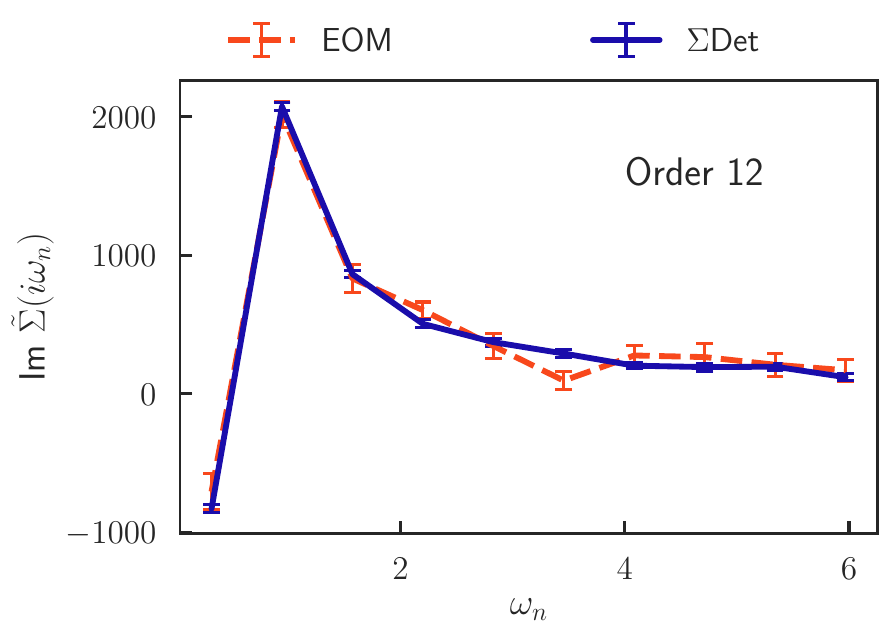}
\vspace{-10pt}
\caption{Imaginary part of the Hubbard atom self-energy
at order 12 in $U$ as obtained from
the equations of motion approach (orange) and the direct self-energy
measurement (blue). We use $\beta = 10$, $U = 1$, $\epsilon=-0.2$. All
simulations lasted 120 CPU hours.}
\label{comp_fbar_atomic}
\end{figure}

\begin{figure}
\centering
\includegraphics[width=0.45\textwidth]{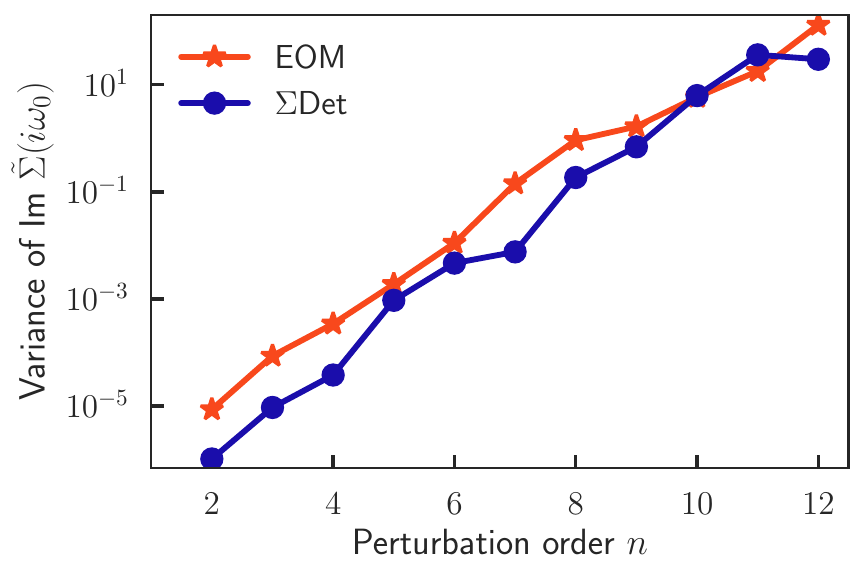}
\vspace{-6pt}
\caption{Variance of the imaginary part of the Hubbard atom self-energy at the
first Matsubara frequency. Orange lines with stars is the result of the
equations of motion. Blue line with dots corresponds to the direct self-energy
measurement.  We use $\beta = 10$, $U = 1$, $\epsilon=-0.2$.}
\label{variance_atomic}
\end{figure}

We now compare the use of equations of motion (EOM) to the direct sampling of
the self-energy expressed as a sum of 1PI diagrams ($\Sigma$Det).  It is not
clear which method is more efficient, as the $\Sigma$Det allows for a precise
cancellation of diagrams and directly samples the quantity of interest but
scales as $n^2 3^n$, while the EOM method cancels diagrams on average but has a
better scaling as $3^n$.

We first consider the Hubbard atom. In Fig.~\ref{comp_fbar_atomic} we show the
contribution to the imaginary part of  the Matsubara frequency
$\tilde{\Sigma}^\sigma(i\omega_n)$ at order 12 for both methods. The equations
of motion method has error bars that are seen to be about 1 order of
magnitude greater than the $\Sigma$Det ones. In order to quantify the
efficiency more accurately, we plot in Fig.~\ref{variance_atomic} the variance
at the first Matsubara frequency $\omega_0$ as a function of the perturbation
order for both methods. We see from this plot that $\Sigma$Det performs better
at low perturbation order, and that both methods tend to become equivalent at
higher orders.

\begin{figure}
\centering
\includegraphics[width=0.5\textwidth]{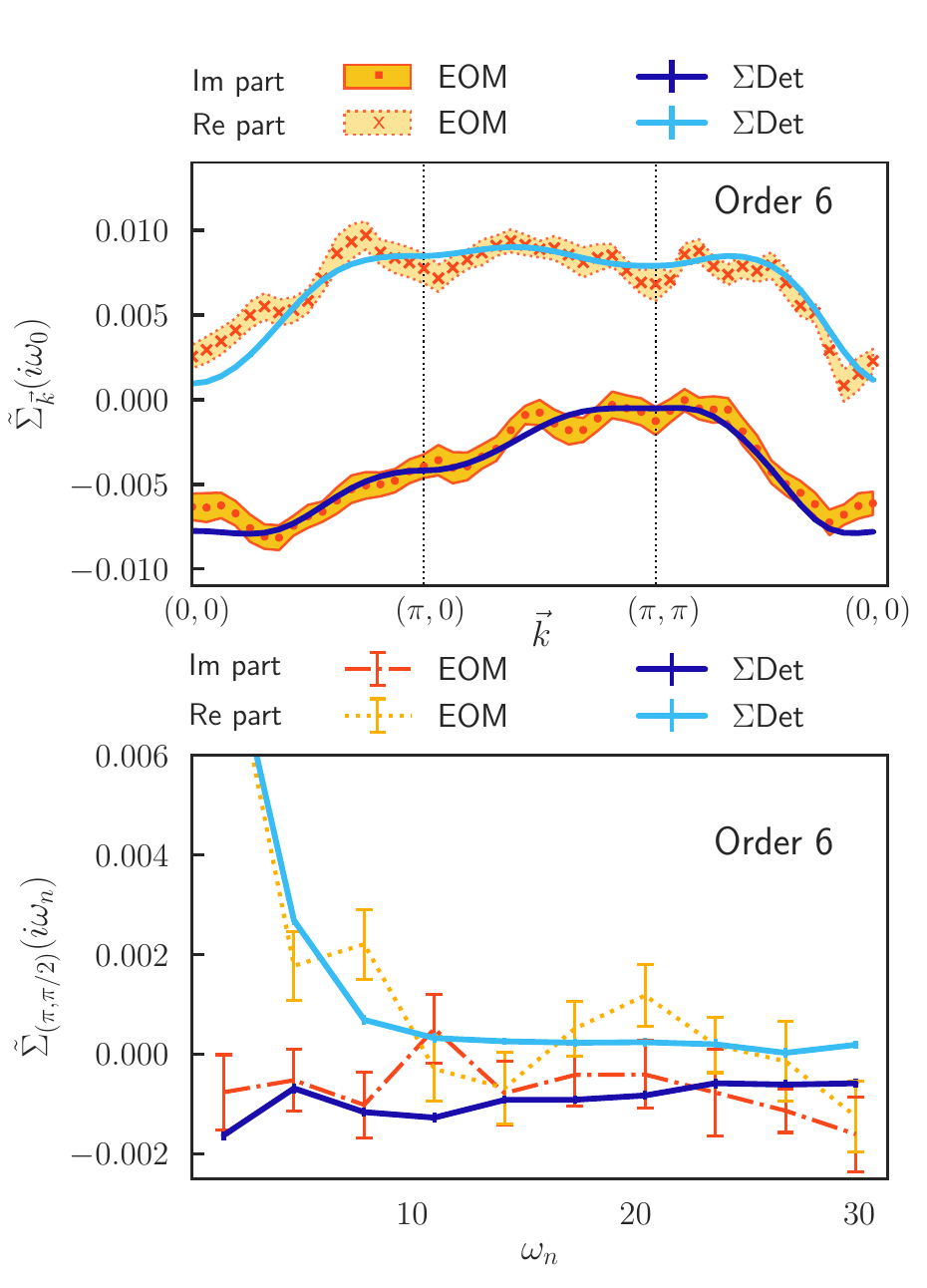}
\vspace{-20pt}
\caption{Hubbard model self-energy at order 6 in $U$ on a $32 \times 32$
lattice with $\beta t = 2$, $U = 4t$, $\mu=0$ and with a uniform $\alpha$ shift
$\alpha_{\uparrow} = \alpha_{\downarrow} = 1.53 t$. Blue symbols are results
for the direct self-energy measurement, orange symbols are results from the
equations of motion approach.  \emph{Upper panel:} Self-energy at the first
Matsubara frequency $\tilde{\Sigma}_{\vec{k}}(i\omega_0)$ along the $\vec{k} =
(0,0) \rightarrow (\pi,0) \rightarrow (\pi,\pi) \rightarrow (0,0)$ path.
\emph{Lower panel:} Self-energy as a function of $i\omega_n$ at $\vec{k} =
(\pi, \pi/2)$. All simulations lasted 120 CPU hours.}
\label{comp_fbar_lattice}
\end{figure}

\begin{figure}
\centering
\includegraphics[width=0.45\textwidth]{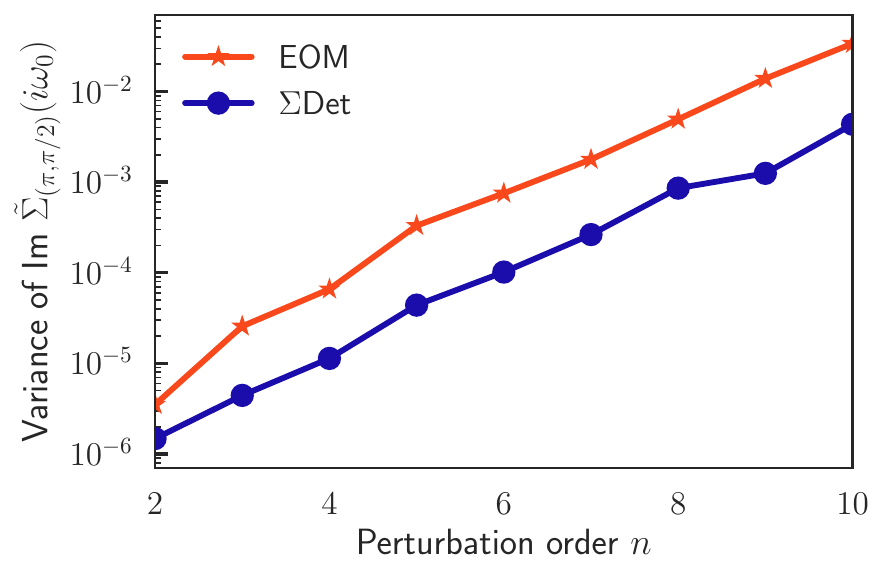}
\vspace{-6pt}
\caption{Variance of the imaginary part of the Hubbard model self-energy Im
$\tilde{\Sigma}_{(\pi,\pi/2)}(i\omega_0)$. Orange lines with stars is the
result equations of motion. Blue line with dots corresponds to the direct
self-energy measurement.  We use a $32 \times 32$ lattice with $\beta t = 2$,
$U = 4t$, $\mu=0$ and with a uniform $\alpha$ shift $\alpha_{\uparrow} =
\alpha_{\downarrow} = 1.53 t$.}
\label{variance_lattice}
\end{figure}

The comparison of the resulting self-energies on the lattice Hubbard model
(Fig.~\ref{comp_fbar_lattice}) shows an even more pronounced  difference
between the two approaches.  At order 6, the contribution to the self-energy
taken at the first Matsubara frequency (upper panel) obtained from $\Sigma$Det
on a chosen path in the Brillouin zone is very well converged and the error
bars for this method are very small (smaller than symbol size). The computation
of $\Sigma^\sigma$ from the equations of motion is less accurate, even if it
agrees with the $\Sigma$Det within its error bars. We then look at the
Matsubara frequency evolution for a given reciprocal lattice vector $\vec{k} =
(\pi,\pi/2)$. The error bar for the EOM method is seen to be large for all
Matsubara frequencies.  To be quantitative, we plot in
Fig.~\ref{variance_lattice} the variance at the first Matsubara frequency
$\omega_0$ for this same value of $\vec{k}=(\pi,\pi/2)$ as function a of the
perturbation order.  We see from this plot that $\Sigma$Det always performs
better than the EOM method, by about one order of magnitude. 

We believe the explanation for this behavior comes from two ingredients.
First, the cancellation of non-one-particle-irreducible diagrams is done
\emph{on average} in the EOM approach, while it is exact in the $\Sigma$Det
algorithm and therefore more efficient to measure the self-energy. This is
particularly visible on the lattice that has more degrees of freedom.  Second,
the self-energy $\Sigma^\sigma$ is more local on the lattice than the
correlator $\bar{F}$. Hence the direct MC sampling of the self-energy still
performs better even though its numerical complexity is greater by a factor $n^2$.
Let us note here that the EOM approach could be useful in the context of the
real-time algorithm of Ref.~\onlinecite{parcollet_prb_2015}. There the
complexity of the EOM approach would be $2^n$ while a direct self-energy
approach would scale as $n^2 3^n$. It may well be that the EOM approach is more
efficient in that case.

\subsection{Comparison between $\Sigma$Det and DiagMC algorithms}

\begin{figure}
\centering
\includegraphics[width=0.51\textwidth]{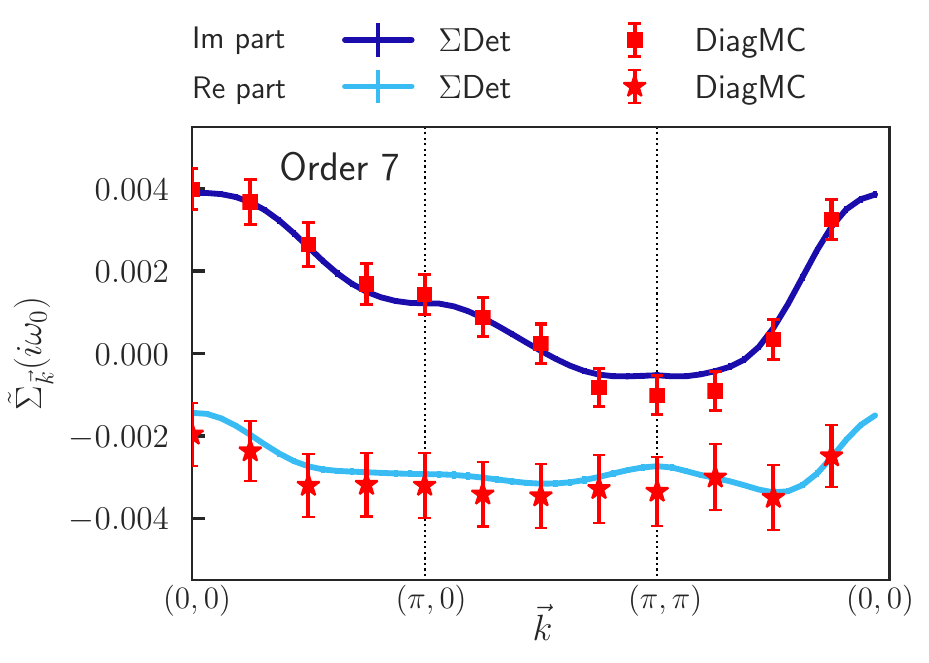}
\vspace{-20pt}
\caption{Hubbard model self-energy at the first Matsubara frequency
$\tilde{\Sigma}_{\vec{k}}(i\omega_0)$ along the $\vec{k} = (0,0) \rightarrow
(\pi,0) \rightarrow (\pi,\pi) \rightarrow (0,0)$ path at order 7 in $U$, as
obtained from DiagMC (red) and the direct self-energy measurement (blue).  We
use a $32 \times 32$ lattice with $\beta t = 2$, $U = 4t$, $\mu=0$ and a
uniform $\alpha$ shift $\alpha_{\uparrow} = \alpha_{\downarrow} = 1.53 t$.
Simulations lasted 1440 CPU hours for the $\Sigma$Det and 4000 CPU hours for
the DiagMC.}
\label{comp_diagmc}
\end{figure}

As the direct calculation of the self-energy $\Sigma$Det proves to be a very
accurate method to get the self-energy, it is natural to compare it to the
state-of-the-art DiagMC results on the two-dimensional Hubbard model. To this
end, we compute in Fig.~\ref{comp_diagmc} the contribution to the first
Matsubara frequency $\omega_0$ of the self-energy at perturbation order 7 for
both $\Sigma$Det and DiagMC methods.  Error bars at this perturbation order,
the highest currently reachable with DiagMC techniques, are much smaller with
the $\Sigma$Det algorithm than with the standard  DiagMC  approach  for
simulations  of  the  same  length.  This algorithm canceling directly non-1PI
diagrams in the MC sampling is therefore an interesting alternative to the
current diagrammatic Monte Carlo approach.

As a final illustration of the method, we compute contributions up to order 9.
The resummed local self-energy is shown in Fig.~\ref{lattice_sigma}.  We
observe that, with a reasonable choice for the $\alpha$ shift, one can
completely converge the results with an uncertainty below $1\%$.

\section{Conclusion}\label{sec:conclusion}

We have introduced and compared three methods to compute the self-energy of
fermionic systems. Two of them rely on the computation of correlators using the
CDet technique, while the third one is an extension of the CDet that allows one to
sum all diagrams that share the same interaction vertices and are one-particle
irreducible.  This allows us to design a Monte Carlo scheme that directly samples
the contributions to the self-energy. This $\Sigma$Det algorithm has an
exponential complexity $n^2 3^n$ where $n$ is the perturbation order.
We have shown that even if it has higher complexity, an approach that computes
the self-energy directly leads to much smaller error bars with respect to the
use of Dyson's equation or more sophisticated equations of motion (nevertheless, the latter
could be useful in the context of real-time quantum Monte Carlo
algorithms).~\cite{parcollet_prb_2015}

\begin{figure}
\centering
\includegraphics[width=0.45\textwidth]{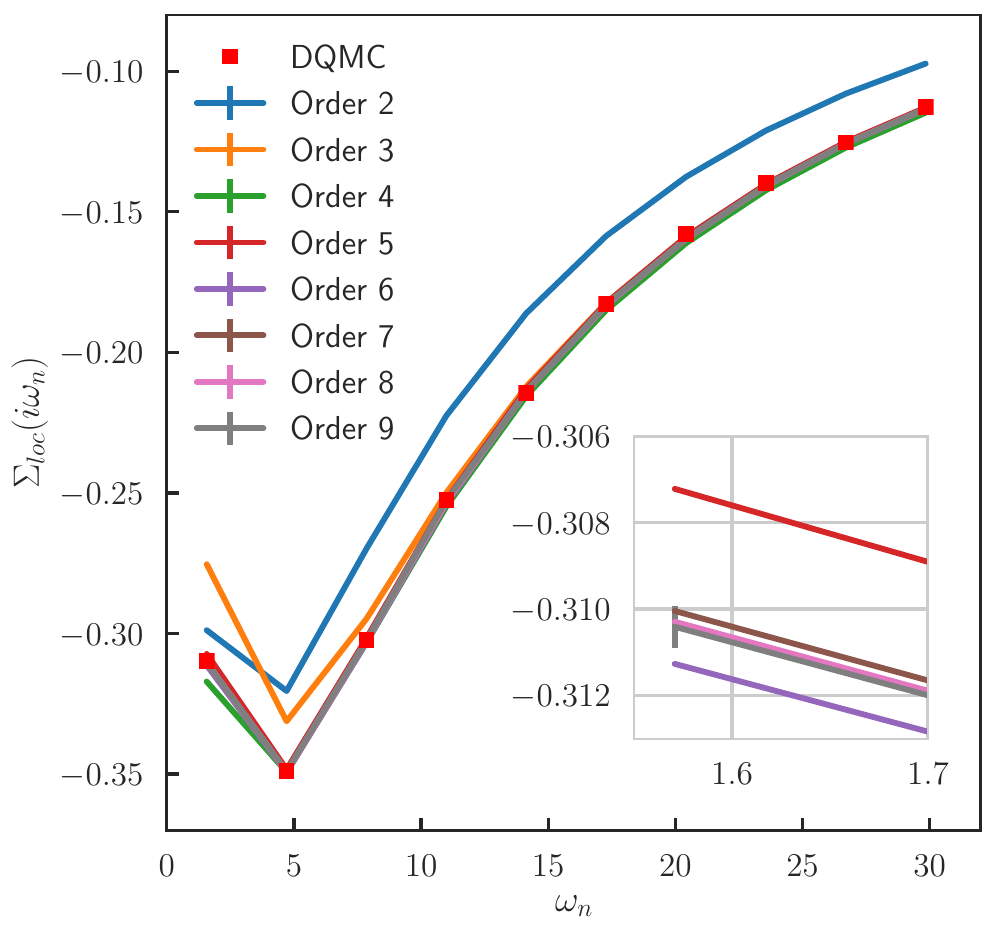}
\vspace{-10pt}
\caption{Imaginary part of the local lattice self-energy
$\Sigma^\sigma_{\text{loc}}(i\omega_n)$ as a function of Matsubara frequency,
as computed using $k$ orders, with $k = 2, \ldots, 9$.  The red squares are
results obtained from DQMC (error bars are smaller than the symbol size).
\emph{Inset:} Zoom on the first Matsubara frequency. It is seen that the
results are converged with an error bar smaller than $1\%$.
We use a $32 \times 32$ lattice with $\beta t = 2, U = 4t, \mu = 0$
and with a uniform $\alpha$ shift $\alpha_\uparrow = \alpha_\downarrow = 1.53t$.
The discrete time interval in DQMC is $\Delta \tau = 1/32$.}
\label{lattice_sigma}
\end{figure}

With the parameters that we have discussed above, $\beta = 2/t$, $U = 4t$ and
$\mu=0$ (corresponding to a total density $n=0.66$), the direct self-energy
measurement also leads to much smaller error bars than the usual DiagMC
algorithm on the two-dimensional Hubbard model and sets the current
state-of-the-art of these approaches. In practice, one can completely converge
the results for 9 orders with an uncertainty below $1\%$.  Note that for these
parameters, other approaches, such as determinant quantum Monte Carlo
(DQMC),~\cite{bss_1981} also converge (see Fig.~\ref{lattice_sigma}). It is
therefore important to more systematically compare the $\Sigma$Det approach,
the DiagMC and other algorithms in different regimes of parameters in order to
determine in what regions of the Hubbard model solutions can be converged. Work
is in progress along these lines (See also the recent article of {\v S}imkovic
and Kozik).~\cite{simkovic_2017}

Finally, further progress is still needed to be able to reach stronger coupling
regimes and lower temperatures. While the summation over all topologies
certainly reduces the sign problem, the stochastic integration over imaginary
times still yields large error bars at high orders. It is therefore necessary
to investigate how this sign problem could be reduced.

\begin{acknowledgments}

We would like to acknowledge valuable feedback from the participants of the
\emph{Diagrammatic Monte Carlo Workshop} held at the Flatiron Institute, June
2017 where this work was first presented and we are particularly grateful for
discussions with A. Georges, E. Kozik, R. Rossi, F. {\v S}imkovic and X.
Waintal. We are especially indebted to O. Parcollet for suggesting improvements
to our work.  This work has been supported by the Simons Foundation within the
Many Electron Collaboration framework (W.W.) and the European Research Council
through Grant No. ERC-319286-QMAC (A.M.). Our codes were developed  using the
TRIQS~\cite{triqs} library. We would also like to thank the computing staff at
CPHT for their help.

\end{acknowledgments}

\begin{appendix}

\section{Equations of motion}\label{app:eom}

Here we show that Eq.~\eqref{eq:sigmafbar} can be obtained from the equations
of motion of the Green's function. For concreteness, we consider the
two-dimensional Hubbard model
\begin{subequations}
\begin{align}
  \mathcal{H} & = -t\sum_{\langle i,j \rangle \sigma} c_{i\sigma}^\dagger c_{j\sigma}
    + U\sum_i n_{i\uparrow}n_{i\downarrow} \\
    & \equiv \mathcal{H}_{\mathrm{hop}} + \mathcal{H}_{\mathrm{int}},
\end{align}
\end{subequations}
where $c_{i\sigma}^\dagger$ creates a spin-$\sigma$ electron on the site $i$ of
a square lattice, $t > 0$ is the nearest-neighbor hopping and $U$ the onsite
interaction. Note that the derivation below yields the same
result for an interacting impurity coupled to a bath or for the Hubbard atom.
These models are used in the article to benchmark and compare results from the
different methods introduced in Sec~\ref{sec:dynamical}.

We define the imaginary-time Green's function of two operators $A$ and $B$ as
$G_{A,B}(\tau) = -\langle T_\tau A(\tau) B(0) \rangle$. The equation of motion
for $G$ is given by
\begin{equation}
  \partial_\tau G_{A,B}(\tau) = -\delta(\tau)\langle \{ A(\tau), B(0) \}\rangle - \langle T_\tau [\mathcal{H},A](\tau)B(0)\rangle,
\end{equation}
which, in Matsubara frequencies, is written
\begin{equation} \label{eq_motion1}
i\omega_n G_{A,B}(i\omega_n) = -G_{[\mathcal{H},A],B}(i\omega_n) + \langle \{A,B\} \rangle.
\end{equation}
Let us note for later use that, by writing $G_{A,B}(\tau) = -\langle T_\tau A(0) B(-\tau) \rangle$,
one obtains a similar expression that involves a commutator between the Hamiltonian and
$B$ rather than $A$
\begin{equation}\label{eq_motion2}
i\omega_n G_{A,B}(i\omega_n) = G_{A,[\mathcal{H},B]}(i\omega_n) + \langle \{A,B\} \rangle.
\end{equation}
The equation of motion (Eq.~\eqref{eq_motion1}) for the one-particle Green's function 
$G_{ij}^\sigma \equiv -\langle T_\tau c_{i\sigma}(\tau)c_{j\sigma}^\dagger(0)\rangle$
is
\begin{equation}
i\omega_n G_{ij}^\sigma = -G_{[\mathcal{H},c_{i\sigma}],c_{j\sigma}^\dagger} + \langle \{c_{i\sigma},c_{j\sigma}^\dagger\}\rangle .
\end{equation}
Using the expression for the commutators
\begin{subequations}
\begin{align}
[\mathcal{H}_{\mathrm{hop}}, c_{i\sigma}] & = t\sum_{\langle a, b\rangle} c_{b\sigma} \delta_{ia}, \\
 [\mathcal{H}_{\mathrm{int}}, c_{i\sigma}] & = -U n_{i\bar{\sigma}}c_{i\sigma},
\end{align}
\end{subequations}
we find that
\begin{subequations}
\begin{align}
i\omega_n G_{ij}^\sigma = -t  \sum_{\langle a, b\rangle} \delta_{ia}G_{bj}^\sigma & +  U G_{n_{i\bar{\sigma}}c_{i\sigma}, c_{j\sigma}^\dagger} + \delta_{ij}, \\
\sum_{\langle a, b \rangle} \left(i\omega_n \delta_{ib} + t\delta_{ia}\right) G_{bj}^\sigma & = U G_{n_{i\bar{\sigma}}c_{i\sigma}, c_{j\sigma}^\dagger} + \delta_{ij}.
\end{align}
\end{subequations}
Introducing the correlator $F_{ij}^\sigma \equiv U
G_{n_{i\bar{\sigma}}c_{i\sigma}, c_{j\sigma}^\dagger}$ the equation above
can be rewritten in matrix form as
\begin{equation}\label{fbarsigmag}
F^\sigma = (G_0^{\sigma -1} - G^{\sigma -1})G^\sigma = \Sigma^\sigma G^\sigma .
\end{equation}
Note that this definition of $F^\sigma$ is consistent with Eq.~\eqref{def_f}.
We can now apply Eq.~\eqref{eq_motion2} to $F_{ij}^\sigma$
\begin{equation}
i\omega_n F_{ij}^\sigma = UG_{n_{i\bar{\sigma}}c_{i\sigma},[\mathcal{H},c_{j\sigma}^\dagger]} + U\left\langle \{ n_{i\bar{\sigma}}c_{i\sigma},c_{j\sigma}^\dagger\}\right\rangle.
\end{equation}
Using the commutators
\begin{subequations}
\begin{align}
[\mathcal{H}_{\mathrm{hop}}, c_{j\sigma}^\dagger] & = -t\sum_{\langle a, b\rangle} c_{a\sigma}^\dagger \delta_{bj}, \\
 [\mathcal{H}_{\mathrm{int}}, c_{j\sigma}^\dagger] & = U n_{j\bar{\sigma}}c_{j\sigma}^\dagger,
\end{align}
\end{subequations}
we find that
\begin{equation}
\sum_{\langle a, b \rangle} \left(i\omega_n \delta_{aj} + t\delta_{bj}\right) F_{ia}^\sigma = U^2 G_{n_{i\bar{\sigma}}c_{i\sigma}, n_{j\bar{\sigma}}c_{j\sigma}^\dagger} + \langle n_{i\bar{\sigma}}\rangle \delta_{ij}.
\end{equation}
Introducing the correlator $\bar{F}_{ij}^\sigma \equiv U^2
G_{n_{i\bar{\sigma}}c_{i\sigma}, n_{j\bar{\sigma}}c_{j\sigma}^\dagger}$ and the
Hartree term $\Sigma_{ij}^{H,\sigma} =  \langle n_{i\bar{\sigma}}\rangle
\delta_{ij}$ the equation above becomes
\begin{equation}
F^\sigma G_0^{\sigma -1} = \bar{F}^\sigma + \Sigma^{H,\sigma}.
\end{equation}
Using Eq.~\eqref{fbarsigmag} for $F^\sigma$ and Dyson's equation we have that
\begin{equation}
F^\sigma G_0^{\sigma -1} = \Sigma^\sigma G^\sigma (G^{\sigma -1} + \Sigma^\sigma) = \Sigma^\sigma + \Sigma^\sigma G^\sigma \Sigma^\sigma,
\end{equation}
which yields the final result
\begin{equation}
\Sigma^\sigma = \Sigma^{H,\sigma} + \bar{F}^\sigma - \Sigma^\sigma G^\sigma \Sigma^\sigma.
\end{equation}
This is the relation between the self-energy and the correlator
$\bar{F}^\sigma$ used in Eq.~\eqref{eq:sigmafbar}. The definitions of
$\bar{F}^\sigma$ and $\Sigma^{H,\sigma}$ are respectively consistent with
Eqs.~\eqref{def_fbar} and~\eqref{hartree_term}.

\section{Benchmarks}\label{sec:bench}

\begin{figure}
\centering
\includegraphics[width=0.45\textwidth]{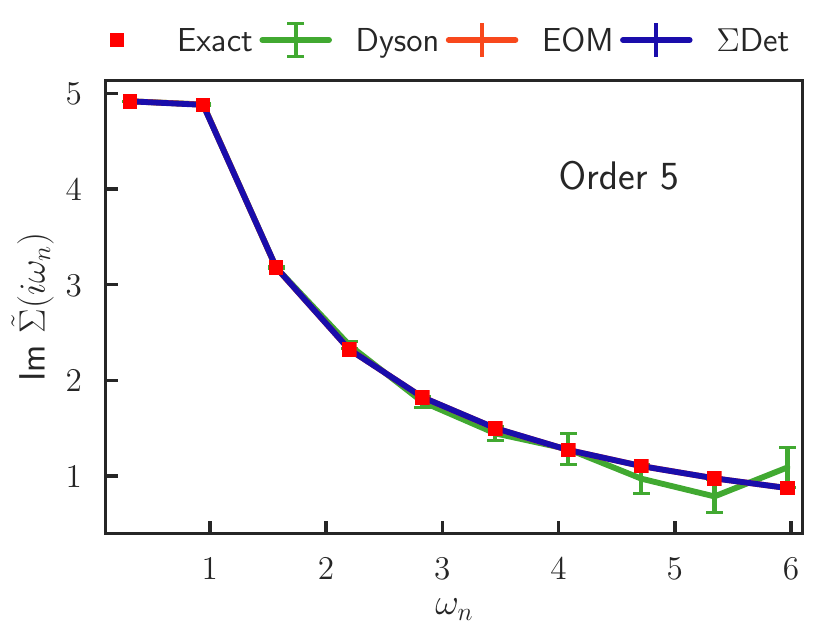}
\vspace{-10pt}
\caption{Benchmark of the contribution to the Matsubara frequency self-energy $\tilde\Sigma(i\omega_n)$ for the Hubbard atom
at order 5 in the perturbation series in $U$. Red squares are the analytical solution. Green lines
are obtained from a calculation of the Green's function with
Eq.~\eqref{rossi_eq}. Orange line is the result of the equations of motion, and lies on top of the blue
curve corresponding to the direct self-energy measurement. We use $\beta = 10$, $U = 1$, $\epsilon=-0.2$. All
simulations lasted 1200 CPU hours.}
\label{benchmark_atomic}
\end{figure}

Here, we present benchmarks for the three methods introduced in the main text
and we check their theoretical complexity.  We first consider the simple
problem of a Hubbard atom.  The self-energy is given by
\begin{equation}
  \Sigma^\sigma(i\omega_n) = \langle n_{\bar{\sigma}} \rangle U
  + \frac{\langle n_{\bar{\sigma}} \rangle (1 - \langle n_{\bar{\sigma}} \rangle) U^2}
         {i\omega_n - \epsilon - (1 - \langle n_{\bar{\sigma}} \rangle) U}
\end{equation}
and the contributions to $\tilde\Sigma(i\omega_n)$ at different orders in $U$ can be
computed analytically. In Fig.~\ref{benchmark_atomic}, we show results for the
contributions to $\tilde\Sigma(i\omega_n)$ at order 5 as obtained from
the proposed algorithms. The results clearly agree with the analytical values within
the error bars.

\begin{figure}
\centering
\includegraphics[width=0.5\textwidth]{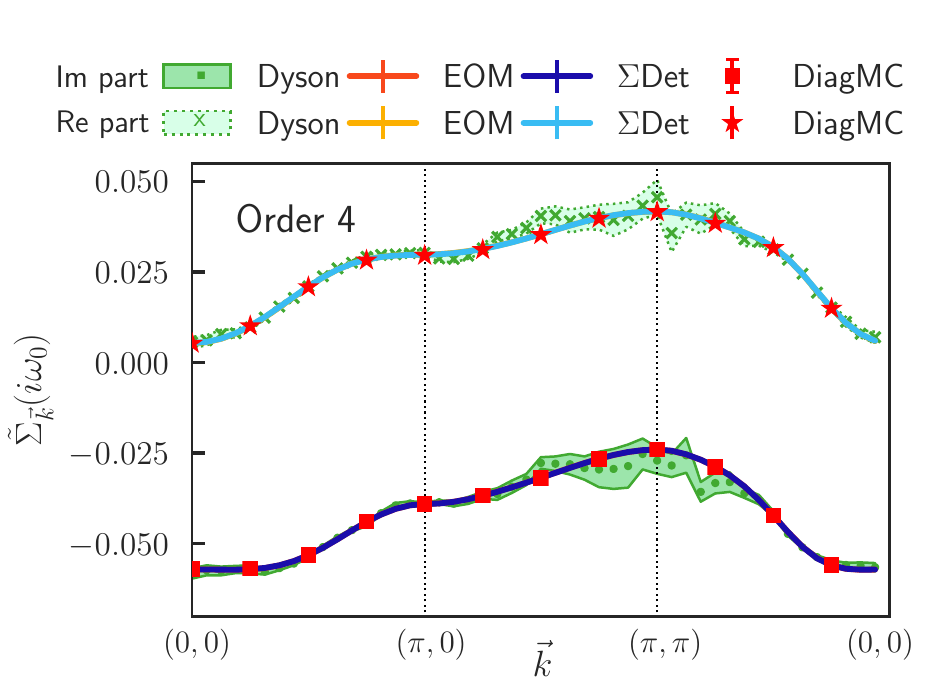}
\vspace{-20pt}
\caption{Hubbard model self-energy at the first Matsubara frequency
$\tilde{\Sigma}_{\vec{k}}(i\omega_0)$ along the $\vec{k} = (0,0) \rightarrow
(\pi,0) \rightarrow (\pi,\pi) \rightarrow (0,0)$ path at order 4 in $U$, as
obtained from DiagMC (red), Dyson's equation (green), the equations of motion approach
(orange) and the direct self-energy measurement (blue). We
use a $32 \times 32$ lattice with $\beta t = 2$, $U = 4t$, $\mu=0$ and a
uniform $\alpha$ shift $\alpha_{\uparrow} = \alpha_{\downarrow} = 1.53 t$.
The DiagMC simulation lasted 400 CPU hours, while all other simulations lasted
1440 CPU hours.}
\label{benchmark_lattice}
\end{figure}

Next we consider the Hubbard model on a $32 \times 32$ square lattice.  In
Fig.~\ref{benchmark_lattice} we plot the momentum-dependent self-energy
$\tilde{\Sigma}_{\vec{k}}(i\omega_0)$ at its first Matsubara frequency along
the $\vec{k} = (0,0) \rightarrow (\pi,0) \rightarrow (\pi,\pi) \rightarrow
(0,0)$ path of the Brillouin zone. Results from the three approaches are are
shown at order 4 and compared to results obtained using the standard
DiagMC~\cite{prokofev_prl_1998, prokofev_prl_2007, kozik_epl_2010,
vanhoucke_2010, burovski_prb_2004} algorithm (This implementation of the
algorithm has been benchmarked and used in earlier calculations, see e.g.
Ref.~\onlinecite{wei_prb_2017}). Results agree with the benchmark DiagMC
calculation within error bars.

\begin{figure}
\centering
\vspace{15pt}
\includegraphics[width=0.4\textwidth]{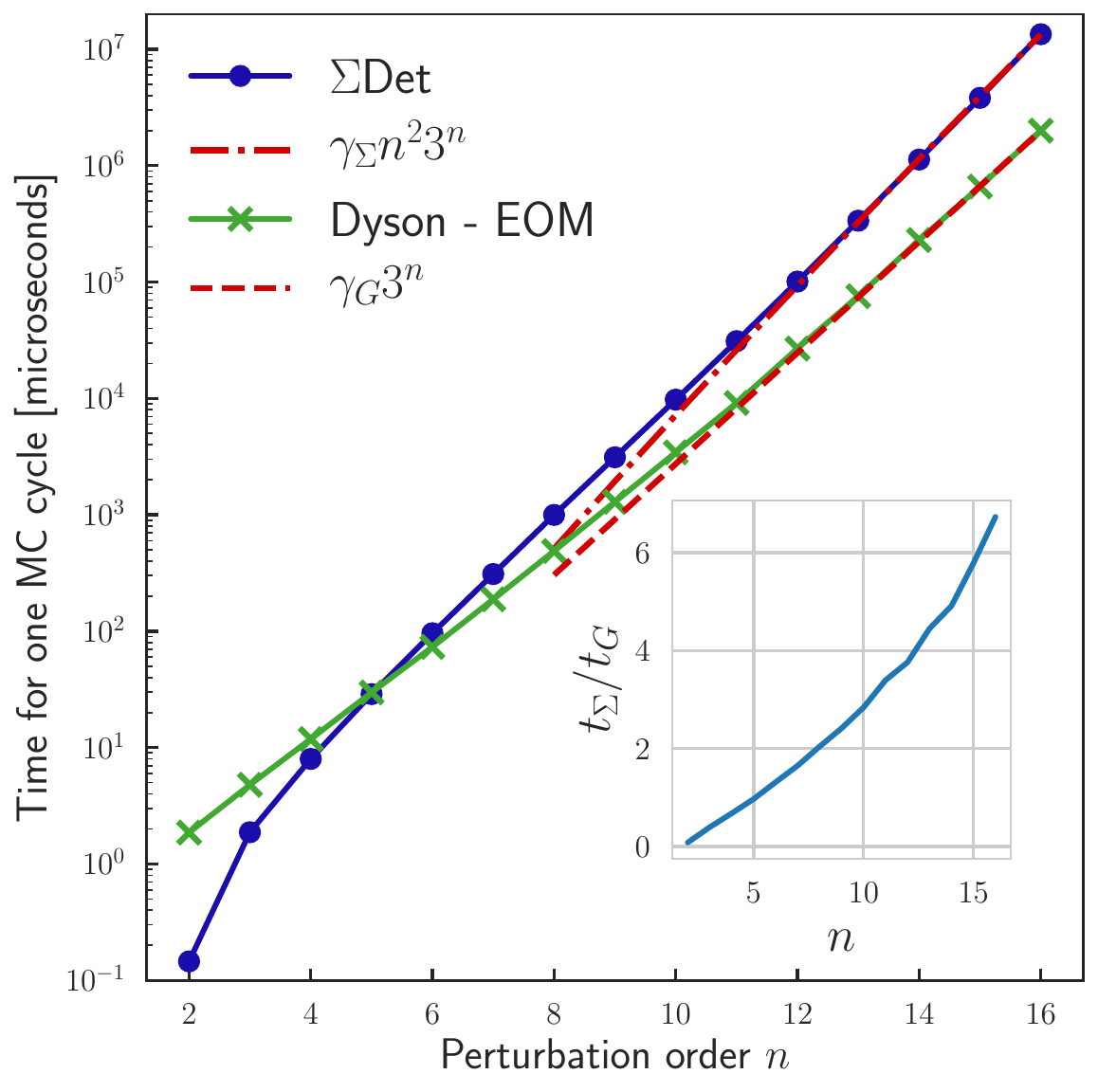}
\vspace{-6pt}
\caption{Comparison of the time for one Monte Carlo cycle (in microseconds)
between the direct accumulation of the self-energy (blue curve with dots) and the computation
of the Green's function using CDet (green curve with dots), on a semilog scale, as a
function of the perturbation order $n$. Each curve is fitted by its expected
high-$n$ behavior: $\gamma_\Sigma n^2 3^n$ for the $\Sigma$Det
(dotted red line) and $\gamma_G 3^n$ for Dyson (dashed red
line), where $\gamma_G = 0.0464$ and $\gamma_\Sigma = 0.0012$ are
implementation-dependent constants.  \emph{Inset}: Ratio of the time of one MC
cycle for the $\Sigma$Det ($t_\Sigma$) and for the CDet
($t_G$), as a function of the perturbation order $n$.}
\label{time_comparison}
\end{figure}

A measurement of the time to perform one MC step allows us to study the complexity
of the algorithms. This is shown in Fig.~\ref{time_comparison}, where the time
for a single step is shown both for the direct measurement of the self-energy
using the $\Sigma$Det and for the measurement of $G$ using the CDet, that is
then used in Dyson's equation.  We know that the EOM method takes twice the
CDet complexity so we consider these two methods together in this study. The
expected high-order behavior in $n^2 3^n$ for the self-energy measurement and
$3^n$ for the CDet is found.
At smaller perturbation orders, the asymptotic behavior is not yet settled. At
orders smaller than 5, the self-energy measurement takes less time mainly
because the algorithm starts at order 2 (The recursion starts with the
pair-bubble diagram, see Eq.~\eqref{start_recursion} with $V = \varnothing$).
On the contrary, the CDet algorithm for the Green's function starts at order 0.
As a consequence, the direct measurement of the self-energy is only about a
factor 3 slower than the CDet approach at order 10 (see inset of
Fig.~\ref{time_comparison}) which is the order that is currently accessible
with reasonable error bars.

\section{Cancellation of non-self-energy diagrams} \label{explicit}

Let us explicitly show the cancellation of non-self-energy diagrams in
Eq.~\eqref{direct_self} for the specific case $V=\{x_1\}$ at order
3 in $U$. We start by considering
\begin{equation}
\begin{fmffile}{appendix1}
\begin{gathered}
\begin{fmfgraph*}(8,8)
\fmfleft{i}
\fmfright{o}
\fmf{plain,right,tension=1}{i,o,i}
\fmf{phantom, label.dist=0,label=$\tilde\Sigma^\sigma_\varnothing$}{i,o}
\fmfdot{i,o}
\fmflabel{$x_{\text{in}}$}{i}
\fmflabel{$x_{\text{out}}$}{o}
\end{fmfgraph*}
\end{gathered}
\hspace{.8cm}
= 
\hspace{.4cm}
\begin{gathered}
\begin{fmfgraph*}(13,8)
\fmfleft{i1,i2}
\fmfright{o1,o2}
\fmf{photon}{i1,i2}
\fmf{photon}{o1,o2}
\fmf{fermion}{i1,o1}
\fmf{fermion,left=0.5,tension=0.2, tag=1}{i2,o2,i2}
\fmfdot{i1}
\fmfdot{o1}
\fmflabel{$x_{\text{in}}$}{i1}
\fmflabel{$x_{\text{out}}$}{o1}
\end{fmfgraph*}
\end{gathered}
\end{fmffile}
\end{equation}
The first term $\bar{F}^\sigma_V(x_{\text{out}}, x_{\text{in}})$ in
Eq.~\eqref{direct_self} corresponds to all connected diagrams with two
external points $\xin$ and $\xout$ and one internal interaction vertex $x_1$:
\begin{fmffile}{appendix2}
\begin{equation}
\begin{split}
& \bar{F}^\sigma_V(x_{\text{out}}, x_{\text{in}}) = 
\begin{gathered}
\begin{fmfgraph*}(15,6)
\fmfleft{i1,i2}
\fmfright{o1,o2}
\fmf{fermion}{i1,v1,o1}
\fmf{phantom}{i2,v2,o2}
\fmf{photon}{i1,i2}
\fmf{photon}{o1,o2}
\fmffreeze
\fmf{fermion,right=0.35,tension=0.5}{i2,o2,i2}
\fmf{phantom,tag=1}{v1,v2}
\fmfipath{p[]}
\fmfiset{p1}{vpath1(__v1,__v2)}
\fmfi{photon}{subpath (0,length(p1)/2) of p1} 
\fmfdot{i1,v1,o1}
\fmflabel{\small $x_{\text{in}}$}{i1}
\fmflabel{\small $x_{\text{out}}$}{o1}
\fmflabel{\small $x_1$}{v1}
\end{fmfgraph*}
\end{gathered}
\hspace{.4cm}
+
\hspace{.4cm}
\begin{gathered}
\begin{fmfgraph*}(15,6)
\fmfleft{i1,i2}
\fmfright{o1,o2}
\fmf{fermion}{i1,v1,o1}
\fmf{phantom}{i2,v2,o2}
\fmf{photon}{i1,i2}
\fmf{photon}{o1,o2}
\fmffreeze
\fmf{fermion,left=0.35,tension=0.5}{i2,o2,i2}
\fmf{phantom,tag=1}{v1,v2}
\fmfipath{p[]}
\fmfiset{p1}{vpath1(__v1,__v2)}
\fmfi{photon}{subpath (0,length(p1)/2) of p1} 
\fmfdot{i1,v1,o1}
\fmflabel{\small $x_{\text{in}}$}{i1}
\fmflabel{\small $x_{\text{out}}$}{o1}
\fmflabel{\small $x_1$}{v1}
\end{fmfgraph*}
\end{gathered} \\ \\
& +
\begin{gathered}
\begin{fmfgraph*}(15,6)
\fmfleft{i1,i2}
\fmfright{o1,o2}
\fmf{fermion}{i1,v1,o1}
\fmf{phantom}{i2,v2,o2}
\fmf{photon}{i1,i2}
\fmf{photon}{o1,o2}
\fmffreeze
\fmf{photon}{v1,v2}
\fmf{fermion,right=0.5,tension=0.5}{i2,v2,i2}
\fmfv{decor.shape=circle,decor.filled=empty, decor.size=.2w}{o2}
\fmfdot{i1,v1,o1}
\fmflabel{\small $x_{\text{in}}$}{i1}
\fmflabel{\small $x_{\text{out}}$}{o1}
\fmflabel{\small $x_1$}{v1}
\end{fmfgraph*}
\end{gathered}
\hspace{.3cm}
+
\hspace{.3cm}
\begin{gathered}
\begin{fmfgraph*}(15,6)
\fmfleft{i1,i2}
\fmfright{o1,o2}
\fmf{fermion}{i1,v1,o1}
\fmf{phantom}{i2,v2,o2}
\fmf{photon}{i1,i2}
\fmf{photon}{o1,o2}
\fmffreeze
\fmf{photon}{v1,v2}
\fmf{fermion,right=0.5,tension=0.5}{v2,o2,v2}
\fmfv{decor.shape=circle,decor.filled=empty, decor.size=.2w}{i2}
\fmfdot{i1,v1,o1}
\fmflabel{\small $x_{\text{in}}$}{i1}
\fmflabel{\small $x_{\text{out}}$}{o1}
\fmflabel{\small $x_1$}{v1}
\end{fmfgraph*}
\end{gathered}
\hspace{.3cm}
+
\hspace{.3cm}
\begin{gathered}
\begin{fmfgraph*}(15,6)
\fmfleft{i1,i2}
\fmfright{o1,o2}
\fmf{fermion}{i1,v1,o1}
\fmf{phantom}{i2,v2,o2}
\fmf{photon}{i1,i2}
\fmf{photon}{o1,o2}
\fmffreeze
\fmf{photon}{v1,v2}
\fmfv{decor.shape=circle,decor.filled=empty, decor.size=.2w}{i2}
\fmfv{decor.shape=circle,decor.filled=empty, decor.size=.2w}{v2}
\fmfv{decor.shape=circle,decor.filled=empty, decor.size=.2w}{o2}
\fmfdot{i1,v1,o1}
\fmflabel{\small $x_{\text{in}}$}{i1}
\fmflabel{\small $x_{\text{out}}$}{o1}
\fmflabel{\small $x_1$}{v1}
\end{fmfgraph*}
\end{gathered}
\end{split}
\end{equation}
\end{fmffile}

\begin{fmffile}{appendix22}
\begin{equation*}
\begin{split}
& + 
\begin{gathered}
\begin{fmfgraph*}(15,15)
\fmfleft{i1,i2}
\fmfright{o1,o2}
\fmf{fermion}{i1,o1}
\fmf{phantom}{i1,v1,i2}
\fmf{photon}{i1,v1}
\fmf{phantom}{o1,v2,o2}
\fmf{photon}{o1,v2}
\fmffreeze
\fmf{fermion,right=0.35,tension=0.5}{v1,v2,v1}
\fmf{phantom}{v1,v3,v2}
\fmf{phantom}{i2,v4,o2}
\fmf{photon,label=\small $x_1$}{v3,v4}
\fmfv{decor.shape=circle,decor.filled=empty, decor.size=.2w}{v4} 
\fmfdot{i1,o1}
\fmflabel{\small $x_{\text{in}}$}{i1}
\fmflabel{\small $x_{\text{out}}$}{o1}
\end{fmfgraph*}
\end{gathered}
\hspace{.3cm}
+
\hspace{.3cm}
\begin{gathered}
\begin{fmfgraph*}(15,15)
\fmfleft{i1,i2}
\fmfright{o1,o2}
\fmf{fermion}{i1,o1}
\fmf{phantom}{i1,v1,i2}
\fmf{photon}{i1,v1}
\fmf{phantom}{o1,v2,o2}
\fmf{photon}{o1,v2}
\fmffreeze
\fmf{fermion,left=0.35,tension=0.5}{v1,v2,v1}
\fmf{phantom}{v1,v3,v2}
\fmf{phantom}{i2,v4,o2}
\fmf{photon,label=\small $x_1$}{v3,v4}
\fmfv{decor.shape=circle,decor.filled=empty, decor.size=.2w}{v4} 
\fmfdot{i1,o1}
\fmflabel{\small $x_{\text{in}}$}{i1}
\fmflabel{\small $x_{\text{out}}$}{o1}
\end{fmfgraph*}
\end{gathered}
\hspace{.3cm}
+
\hspace{.3cm}
\begin{gathered}
\begin{fmfgraph*}(15,10)
\fmfleft{i1,i2}
\fmfright{o1,o2}
\fmf{fermion}{i1,v1,o1}
\fmf{photon}{i1,v2,i2}
\fmf{photon}{o1,v3,o2}
\fmf{fermion,right=0.35,tension=0.5}{i2,o2,i2}
\fmffreeze
\fmf{phantom}{v2,v4,v3}
\fmf{photon}{v1,v4}
\fmfv{decor.shape=circle,decor.filled=empty, decor.size=.2w}{v4}
\fmfdot{i1,v1,o1}
\fmflabel{\small $x_{\text{in}}$}{i1}
\fmflabel{\small $x_{\text{out}}$}{o1}
\fmflabel{\small $x_1$}{v1}
\end{fmfgraph*}
\end{gathered} \\ \\
& \hspace{1.8cm} + 
\begin{gathered}
\begin{fmfgraph*}(10,13)
\fmfleft{i1,i2}
\fmfright{o1,o2}
\fmf{fermion}{i1,o1}
\fmf{phantom}{i1,v1,i2}
\fmf{phantom}{o1,v2,o2}
\fmffreeze
\fmf{photon}{i1,v1}
\fmfv{decor.shape=circle,decor.filled=empty, decor.size=.2w}{v1}
\fmf{photon}{o1,v2}
\fmfv{decor.shape=circle,decor.filled=empty, decor.size=.2w}{v2}
\fmf{photon,label=\small $x_1$}{v2,o2}
\fmfv{decor.shape=circle,decor.filled=empty, decor.size=.2w}{o2}
\fmfdot{i1,o1}
\fmflabel{\small $x_{\text{in}}$}{i1}
\fmflabel{\small $x_{\text{out}}$}{o1}
\end{fmfgraph*}
\end{gathered}
\hspace{.3cm}
+
\hspace{.3cm}
\begin{gathered}
\begin{fmfgraph*}(10,13)
\fmfleft{i1,i2}
\fmfright{o1,o2}
\fmf{fermion}{i1,o1}
\fmf{phantom}{i1,v1,i2}
\fmf{phantom}{o1,v2,o2}
\fmffreeze
\fmf{photon}{i1,v1}
\fmfv{decor.shape=circle,decor.filled=empty, decor.size=.2w}{v1}
\fmf{photon,label=\small $x_1$}{v1,i2}
\fmfv{decor.shape=circle,decor.filled=empty, decor.size=.2w}{i2}
\fmf{photon}{o1,v2}
\fmfv{decor.shape=circle,decor.filled=empty, decor.size=.2w}{v2}
\fmfdot{i1,o1}
\fmflabel{\small $x_{\text{in}}$}{i1}
\fmflabel{\small $x_{\text{out}}$}{o1}
\end{fmfgraph*}
\end{gathered} 
\end{split}
\end{equation*}
\end{fmffile}
From this sum, we subtract the second and third terms of Eq.~\eqref{direct_self}. The former gives
\begin{equation}
\begin{fmffile}{appendix3}
(\Sigma^\sigma G^\sigma)_\varnothing(x_{\text{out}}, x_1)\tilde\Sigma^\sigma_\varnothing(x_1,x_{\text{in}}) =
\begin{gathered}
\begin{fmfgraph*}(15,6)
\fmfleft{i1,i2}
\fmfright{o1,o2}
\fmf{fermion}{i1,v1,o1}
\fmf{phantom}{i2,v2,o2}
\fmf{photon}{i1,i2}
\fmf{photon}{o1,o2}
\fmffreeze
\fmf{photon}{v1,v2}
\fmf{fermion,right=0.5,tension=0.5}{i2,v2,i2}
\fmfv{decor.shape=circle,decor.filled=empty, decor.size=.2w}{o2}
\fmfdot{i1,v1,o1}
\fmflabel{\small $x_{\text{in}}$}{i1}
\fmflabel{\small $x_{\text{out}}$}{o1}
\fmflabel{\small $x_1$}{v1}
\end{fmfgraph*}
\end{gathered}
\end{fmffile}
\vspace{0.4cm}
\end{equation}
while the latter's contribution is the sum of
\begin{equation}
\begin{fmffile}{appendix4}
F^\sigma_\varnothing(x_{\text{out}}, x_{\text{in}}) \left(UG^{\bar{\sigma}}_{\{x_1\}}(0^-)\right)  = 
\begin{gathered}
\begin{fmfgraph*}(10,13)
\fmfleft{i1,i2}
\fmfright{o1,o2}
\fmf{fermion}{i1,o1}
\fmf{phantom}{i1,v1,i2}
\fmf{phantom}{o1,v2,o2}
\fmffreeze
\fmf{photon}{i1,v1}
\fmfv{decor.shape=circle,decor.filled=empty, decor.size=.2w}{v1}
\fmf{photon,label=\small $x_1$}{v1,i2}
\fmfv{decor.shape=circle,decor.filled=empty, decor.size=.2w}{i2}
\fmf{photon}{o1,v2}
\fmfv{decor.shape=circle,decor.filled=empty, decor.size=.2w}{v2}
\fmfdot{i1,o1}
\fmflabel{\small $x_{\text{in}}$}{i1}
\fmflabel{\small $x_{\text{out}}$}{o1}
\end{fmfgraph*}
\end{gathered}
\end{fmffile}
\end{equation}
and of
\begin{fmffile}{appendix5}
\begin{equation}
\begin{split}
F^\sigma_{\{ t_1\}} & (x_{\text{out}}, x_{\text{in}})  \left(UG^{\bar{\sigma}}_\varnothing(0^-)\right) =
\hspace{.3cm}
\begin{gathered}
\begin{fmfgraph*}(10,13)
\fmfleft{i1,i2}
\fmfright{o1,o2}
\fmf{fermion}{i1,o1}
\fmf{phantom}{i1,v1,i2}
\fmf{phantom}{o1,v2,o2}
\fmffreeze
\fmf{photon}{i1,v1}
\fmfv{decor.shape=circle,decor.filled=empty, decor.size=.2w}{v1}
\fmf{photon}{o1,v2}
\fmfv{decor.shape=circle,decor.filled=empty, decor.size=.2w}{v2}
\fmf{photon,label=\small $x_1$}{v2,o2}
\fmfv{decor.shape=circle,decor.filled=empty, decor.size=.2w}{o2}
\fmfdot{i1,o1}
\fmflabel{\small $x_{\text{in}}$}{i1}
\fmflabel{\small $x_{\text{out}}$}{o1}
\end{fmfgraph*}
\end{gathered} \\ \\
& +
\begin{gathered}
\begin{fmfgraph*}(15,6)
\fmfleft{i1,i2}
\fmfright{o1,o2}
\fmf{fermion}{i1,v1,o1}
\fmf{phantom}{i2,v2,o2}
\fmf{photon}{i1,i2}
\fmf{photon}{o1,o2}
\fmffreeze
\fmf{photon}{v1,v2}
\fmf{fermion,right=0.5,tension=0.5}{v2,o2,v2}
\fmfv{decor.shape=circle,decor.filled=empty, decor.size=.2w}{i2}
\fmfdot{i1,v1,o1}
\fmflabel{\small $x_{\text{in}}$}{i1}
\fmflabel{\small $x_{\text{out}}$}{o1}
\fmflabel{\small $x_1$}{v1}
\end{fmfgraph*}
\end{gathered}
\hspace{.4cm}
+
\hspace{.4cm}
\begin{gathered}
\begin{fmfgraph*}(15,6)
\fmfleft{i1,i2}
\fmfright{o1,o2}
\fmf{fermion}{i1,v1,o1}
\fmf{phantom}{i2,v2,o2}
\fmf{photon}{i1,i2}
\fmf{photon}{o1,o2}
\fmffreeze
\fmf{photon}{v1,v2}
\fmfv{decor.shape=circle,decor.filled=empty, decor.size=.2w}{i2}
\fmfv{decor.shape=circle,decor.filled=empty, decor.size=.2w}{v2}
\fmfv{decor.shape=circle,decor.filled=empty, decor.size=.2w}{o2}
\fmfdot{i1,v1,o1}
\fmflabel{\small $x_{\text{in}}$}{i1}
\fmflabel{\small $x_{\text{out}}$}{o1}
\fmflabel{\small $x_1$}{v1}
\end{fmfgraph*}
\end{gathered}
\end{split}
\end{equation}
\end{fmffile}
We see that the remaining contributions to the self-energy that remain are only
those diagrams that are one-particle irreducible.

\end{appendix}

\bibliography{dynamical_det}

\end{document}